\pdfoutput=1
\documentclass[sigconf,nonacm]{acmart}
\usepackage{enumitem}
\usepackage{dblfloatfix}
\usepackage{subcaption}
\usepackage{wrapfig}
\usepackage{quoting}
\usepackage{amsmath}
\usepackage{booktabs}
\usepackage{array}
\usepackage{graphicx}

\makeatletter\newcommand{\qmark}{?\@ifnextchar.{\@gobble}{}}\makeatother  %

\AtBeginDocument{%
}

\begin{document}

\title[Who are \textit{you?} The Personal Imitation Game]{Who are \textit{you?} The Personal Imitation Game}

\author{Muhammad Rizky Wellyanto, Hajara-Yasmin Isa, Kexin Xue, Vishal Ravichandran, Wenpeng Lu, Jerry O. Talton, Ranjitha Kumar}
\affiliation{%
  \department{Department of Computer Science}
  \institution{University of Illinois at Urbana-Champaign}
  \city{Urbana}
  \state{Illinois}
  \country{USA}
}
\email{{wellyan2, hisa2, kexinx4, vishal8, wenpeng6, talton, ranjitha}@illinois.edu}
\renewcommand{\shortauthors}{Wellyanto et al.}
\begin{abstract}
Large Language Models (LLMs) now perform strongly on Alan Turing's 1950 Imitation Game, so we propose a harder challenge --- the \textit{Personal} Imitation Game --- which asks whether a machine can replicate a \textit{specific} individual in conversation rather than a generic human. Testing with 30 participants, we establish a human baseline and report the identity signals interrogators rely on: stylometric features, personality traits, and biographical facts. To demonstrate the framework, we design an LLM-based system that plays the game for one of the paper's authors, drawing on an autoethnographic dataset spanning ten years of personal digital communication. In a 20-person deception study, interrogators choose our system over the author's own friends in 7 of 10 rounds, and the author over our system in 13 of 20; neither rate is distinguishable from chance at this sample size. We consider design improvements, applications of personal AI agents, and implications for identity and privacy.
\end{abstract}

\ccsdesc[500]{Human-centered computing~Natural language interfaces}
\ccsdesc[300]{Human-centered computing~Empirical studies in HCI}
\ccsdesc[300]{Computing methodologies~Natural language generation}
\keywords{Personal generative agents, autoethnography, LLMs}

\begin{teaserfigure}
	\centering
	\vspace{-.11in}
	\includegraphics[width=\textwidth]{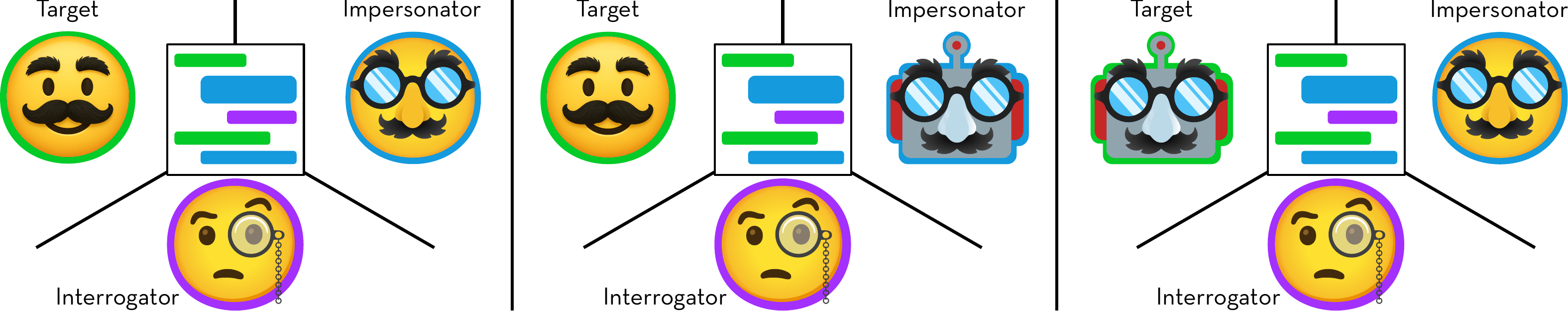}
	\caption{Three variants of the Personal Imitation Game. In each one, a
	human \textit{interrogator} chats with two anonymized participants:
	a \textit{target} --- whose identity is known --- and an \textit
	{impersonator} who attempts to mimic the target's identity.  In the
	first variant (left), both the target and impersonator are humans, and
	all three participants are known to one another.  In the second
	(center), the target is one of the paper's authors, the interrogator is
	a (human) friend or colleague, and the impersonator is an LLM-based
	system powered by the author's autoethnography, which includes ten years
	of personal message data.  In the third (right), the LLM takes the
	target author's place, and the impersonator is again a friend.}
	\label{fig:pig}
	\Description{Three horizontally arranged panels depict variants of the Personal Imitation Game. In each panel, three roles—Interrogator, Target, and Impersonator—are positioned radially around a central chat interface. Thin radial lines form pizza-slice separators between roles, indicating physical separation and text-only communication. Role labels and icons are present: a monocled thinking-face for the human Interrogator; a mustached human for a real human Target; a human with a fake mustache for a human Impersonator; and a robot icon for AI roles (the robot also wears a fake mustache when acting as the Impersonator). Chat bubbles are color-coded with a consistent mapping across panels—blue for Interrogator, green for Target, purple for Impersonator—supplemented by role labels and matching icon halos to avoid reliance on color alone. Left panel: all three roles are humans (all-human baseline). Center panel: human Target (the author), human Interrogator (friend/colleague), and an AI Impersonator grounded in the author's data. Right panel: the Target is an AI model standing in for the author, and the Impersonator is a human. The figure foregrounds text-only, turn-taking interaction and a controlled comparison across variants.}
\end{teaserfigure}

\maketitle

\section{Introduction}

\begin{samepage}
\begin{quoting}[leftmargin=0.05\columnwidth]
	``Who are \textit{you?}'' said the Caterpillar.\\ This was not an encouraging opening for a conversation.
	\begin{flushright}
		\textit{Alice's Adventures in Wonderland}, Lewis Carroll
	\end{flushright}
\end{quoting}
\end{samepage}
\noindent In 1950, Alan Turing asked: ``Can machines
 think?'' To answer it, he proposed a challenge he called the Imitation Game,
 now widely known as the Turing Test~\cite{computing_turing_1950}. In the
 test, a human interrogator exchanges written messages with another human and a
 computer ``pretending'' to be human: the game is won if the computer is able
 to fool the interrogator. Despite decades of fruitful work in Artificial
 Intelligence and Machine Learning after Turing proposed his test, a system
 capable of winning the Imitation Game remained stubbornly out of reach for
 nearly seventy years.

Transformer-based neural networks~\cite
{attention_vaswani_2017} shifted this status quo within a few years~\cite
{jones2025largelanguagemodelspass,human_or_not_jannai_2023,Biever2023ChatGPTBrokeTuringTest}.
Large Language Models (LLMs) like GPT-3.5~\cite
{language_brown_2020,training_ouyang_2022}
quickly demonstrated remarkable proficiency in ``understanding'' and
emulating human conversation~\cite{openai2024gpt4technicalreport,llama_touvron_2023,geminiteam2024geminifamilyhighlycapable}. These foundation models particularly
excel at synthesizing their vast training repositories, enabling a number of
general-purpose applications like search engines that summarize their
results, question-and-answer software interfaces, and chat agents that can
perform complex tasks (e.g., plan a multi-city trip)~\cite
{language_brown_2020, bommasani2022opportunitiesrisksfoundationmodels}.

This rapid progress on generalized conversational agents opens the door to a
new, exciting class of capabilities: the development of \textit
{personal} language models (sometimes called digital ``clones'' or ``twins''). Imagine a
virtual chief of staff that responds to your emails as you would, with
recipients none the wiser; a dating concierge that simulates thousands of
first dates, saving you time and emotional labor; or a chat interface that
your grandchildren can use to seek your advice long after you have passed
away~\cite{brubaker_ai_afterlife_2024}. A growing body of work --- exemplified by OpenClaw~\cite{openclaw2025, openclawsite2025} and PAHF~\cite{liang2025pahf} --- pursues what is often called \textit{personalized} AI: agents that act \textit{on behalf of} a user, adapting to their preferences and context. We draw a deliberate distinction: \textit{personalized} agents are tools customized \textit{for} you; \textit{personal} agents replicate \textit{you} --- your voice, your reasoning, your identity in conversation. This paper is concerned with the latter. Turing asked, ``Can machines
think?''; this paper considers the question ``can a machine think \textit
{like me}?''

To answer this question, we propose the \textit{Personal} Imitation Game
(PIG), a new variant of the Turing Test. In the game, a human \textit
{interrogator} engages in written dialogue with two participants --- a \textit{target} and an \textit{impersonator} --- all known to each
other. The participants are given a conversational prompt and instructed to
discuss it; the target is told to converse naturally, while the impersonator
is charged with imitating the target as convincingly as possible. The
interrogator wins the game if they successfully determine which participant
is the real target at the end of the chat (Figure~\ref{fig:pig}).

We establish a baseline for the game by playing it with 10 groups of 3 human
participants, and report qualitative and quantitative analyses of this study:
how successfully people can discern identity from text, the features they use
to make judgments, how well they assess their own uncertainty, and how the
strength of their interpersonal relationships correlates with success and
decision-making. Then, we implement an LLM-based system leveraging the
signals observed in our human baseline --- stylometric features
(e.g., vocabulary, word choice), personality traits (e.g., signs of
curiosity, indications of diligence), and biographical facts (e.g., age,
gender) --- to play the PIG for one of the paper's authors, drawing on an
autoethnography that includes a decade's worth of personal chat
messages~\cite{ellis-authoethnography-10}. We evaluate this system via a
deception study involving 20 of the author's friends and colleagues. Interrogators chose
the system over the author's friends in 7 of 10 rounds, and chose the author
over the system in 13 of 20 rounds. With 10 and 20 rounds our study is powered
only for large effects, and neither rate differs from chance (exact
binomial $p = 0.34$ and $p = 0.26$), so we treat both as suggestive point
estimates and rest our contribution on the framework rather than on the
magnitude of these rates.

Our results sketch the early promise of personal chat agents, and
motivate exploring alternative approaches in larger studies. Together, the framework and its proof-of-concept
evaluation offer a foundation for future work on personal AI agents. We
sketch promising future directions for improving personal language
models~\cite{ling2024domainspecializationkeymake,
beltagy2019scibertpretrainedlanguagemodel}, speculate on the rich space of
applications they may enable, and discuss the implications for digital
identity and personal privacy. These implications are no longer
speculative: commercial assistants now connect to personal message archives,
and the strongest public objection to one such release spoke for the
conversation partners who never agreed to the
connection~\cite{bloomberg2026chatgpt_imessage, moraco2026tweet,
hoag2026tweet}.
\section{Related Work}

Since 2023, large language
models have performed strongly on several Turing Test-style
evaluations~\cite
{jones2025largelanguagemodelspass,
	gpt4_jones_2023, human_or_not_jannai_2023}. Newer systems aim to construct
LLMs to embody specific personas~\cite
{park2023generativeagentsinteractivesimulacra,
	zhang2018personalizingdialogueagents_personachat}, or even particular
individuals~\cite
{park2024generativeagentsimulations1000,
	toubia2025twin2k500datasetbuildingdigital, herbold2024large}. To assess the
fidelity of such personalized models, most prior work has focused on \textit
{alignment}: measuring the extent to which a model's survey responses or
game-theoretic decisions mirror a human ground-truth~\cite
{park2024generativeagentsimulations1000,
	toubia2025twin2k500datasetbuildingdigital}.
\subsection{The Imitation Game \& LLMs}

Alan Turing introduced the Imitation Game --- now commonly known as the Turing
Test --- as an operational test for conversational intelligence~\cite
{computing_turing_1950}. In the game, a human interrogator exchanges text
messages with two hidden participants: a woman (the target) and a man
instructed to impersonate her. The interrogator's task is to decide which
participant is the woman. Turing proposed replacing the male impersonator
with a machine and asked whether the interrogator would decide wrongly as
often as in the man--woman version. (His five-minute, 70\% figure
belongs to a separate prediction about machines circa the year 2000, not to
the criterion itself.)
Researchers have long treated the Turing Test as a benchmark for evaluating AI systems'
human-likeness. Early conversational programs were read as candidates for it ---
ELIZA~\cite{eliza_weizenbaum96} inadvertently, as Weizenbaum himself
objected, and PARRY~\cite{colby_parry_1972} through explicit
Turing-like indistinguishability tests. For decades, however, no system
convincingly succeeded, prompting critiques of the test's anthropocentric
assumptions~\cite{turing_french_2000, turing_saygin_2000}, its adequacy as
a measure of understanding~\cite{minds_searle_1988}, its reliance on deception~\cite
{lessons_shieber_1994}, and its status as a philosophical
thought experiment~\cite{goncalves_tt_thought_exp_2023}. As a consequence,
much evaluation of AI linguistic capabilities shifted toward narrower,
task-specific benchmarks --- such as summarization~\cite
{Nenkova2012ASO}, part-of-speech tagging~\cite
{marcus-etal-1993-building}, and machine translation~\cite
{koehn2009statistical} --- where performance can be measured more precisely.

The emergence of LLMs such as GPT~\cite
{openai2024gpt4technicalreport, language_brown_2020}, LLaMA~\cite
{llama_touvron_2023}, and Gemini~\cite
{geminiteam2024geminifamilyhighlycapable} has revived interest in the Turing
Test~\cite
{Biever2023ChatGPTBrokeTuringTest, jones2025largelanguagemodelspass,
gpt4_jones_2023, human_or_not_jannai_2023}. Jones and Bergen~\cite
{jones2025largelanguagemodelspass, gpt4_jones_2023} show that carefully
prompted LLMs can be difficult to distinguish from humans in the Imitation
Game, though the effect is strongly prompt-dependent: GPT-4.5 with a persona
prompt was judged human in 73\% of games, while their best GPT-4 prompt
passed in only 49.7\%, short of the 66\% baseline set by human participants. Jannai et al.~\cite{human_or_not_jannai_2023} ran a
large-scale two-party variant in which over 1.5 million players had
two-minute chats with either a human or one of several persona-prompted LLMs;
players identified their partner correctly in 68\% of games overall, and in
only 60\% of games against a bot. These
studies demonstrate that LLMs can convincingly masquerade as a human in
conversation.

\subsection{Personas, Twins, \& Clones}

Building on an LLM's ability to generate human-like responses, newer systems
aim to simulate \textit{personas} --- or representations of specific
sub-populations --- by assigning LLM chat agents synthetic demographics,
occupations, and biographies~\cite
{park2023generativeagentsinteractivesimulacra,
	samuel2024personagymevaluatingpersonaagents}. Persona-conditioned agents
can respond more appropriately to human emotions and social cues~\cite
{ma_exploring_2019},
as well as improving response quality and engagement during
interactions~\cite
{tseng2024talespersonallmssurvey,
zhang2018personalizingdialogueagents_personachat, towards_adiwardana_2020,
Salemi2023}. Moreover, assembling collections of persona-based agents enables
the simulation of population-level behaviors~\cite
{park2023generativeagentsinteractivesimulacra}.

Taking persona simulation to an individual level has led to a number of
efforts to create AI \textit{clones}, or \textit{digital twins}\footnote
{Despite its apt application to personalized AI, the term ``digital twin'' long
	predates LLMs as a term of art in physical design and manufacturing~\cite
	{grieves2016digitaltwins}}: agents that impersonate \textit
{specific} individuals. Clones and twins are typically constructed by
providing LLMs with personal data collected via interviews~\cite
{park2024generativeagentsimulations1000}, public speech or writing~\cite
{herbold2024large}, or surveys~\cite
{toubia2025twin2k500datasetbuildingdigital}. To evaluate the fidelity of the
resultant models, some authors have examined \textit{alignment} between an
individual and their LLM counterpart using standardized surveys (e.g., the
General Social Survey~\cite{GSS2024}, Big Five personality scales~\cite
{John1999_big_five}) and behaviors in economic games~\cite
{park2024generativeagentsimulations1000,
	toubia2025twin2k500datasetbuildingdigital}.
Later work stress-tests this paradigm at scale. Across 19
pre-registered studies, Peng et al.~\cite{peng2025funhousemirrors} report that
twins built from more than 500 self-report answers per
participant~\cite{toubia2025twin2k500datasetbuildingdigital} correlate only
weakly with that participant's own responses ($r = 0.20$ across 164 outcomes),
and beat a base LLM given no individual data only modestly. Peng et al. name
five systematic distortions; the first, insufficient individuation, is the
failure the Personal Imitation Game targets:
a model can match a population while missing the person.

Two efforts evaluate individual-level impersonation directly and are the
closest prior work to ours. IMPersona~\cite{shi2025impersona} grounds a model in
an individual's own message history and asks judges who know that individual
whether a single chat partner is human. The Individual Turing
Test~\cite{guo2026individualturingtest} draws on a volunteer's decade-long
private messaging archive and asks acquaintances to rank candidate replies by
how plausibly each came from the target, finding that current methods fail
against acquaintances while performing far better against strangers.
IMPersona finds that fine-tuning outperforms prompting, whereas the
Individual Turing Test reports a trade-off: fine-tuning better captures
everyday chat style, but retrieval- and memory-based methods perform better
on personal opinions and preferences.

\begin{figure*}[t]
	\centering
	\includegraphics[width=\textwidth]{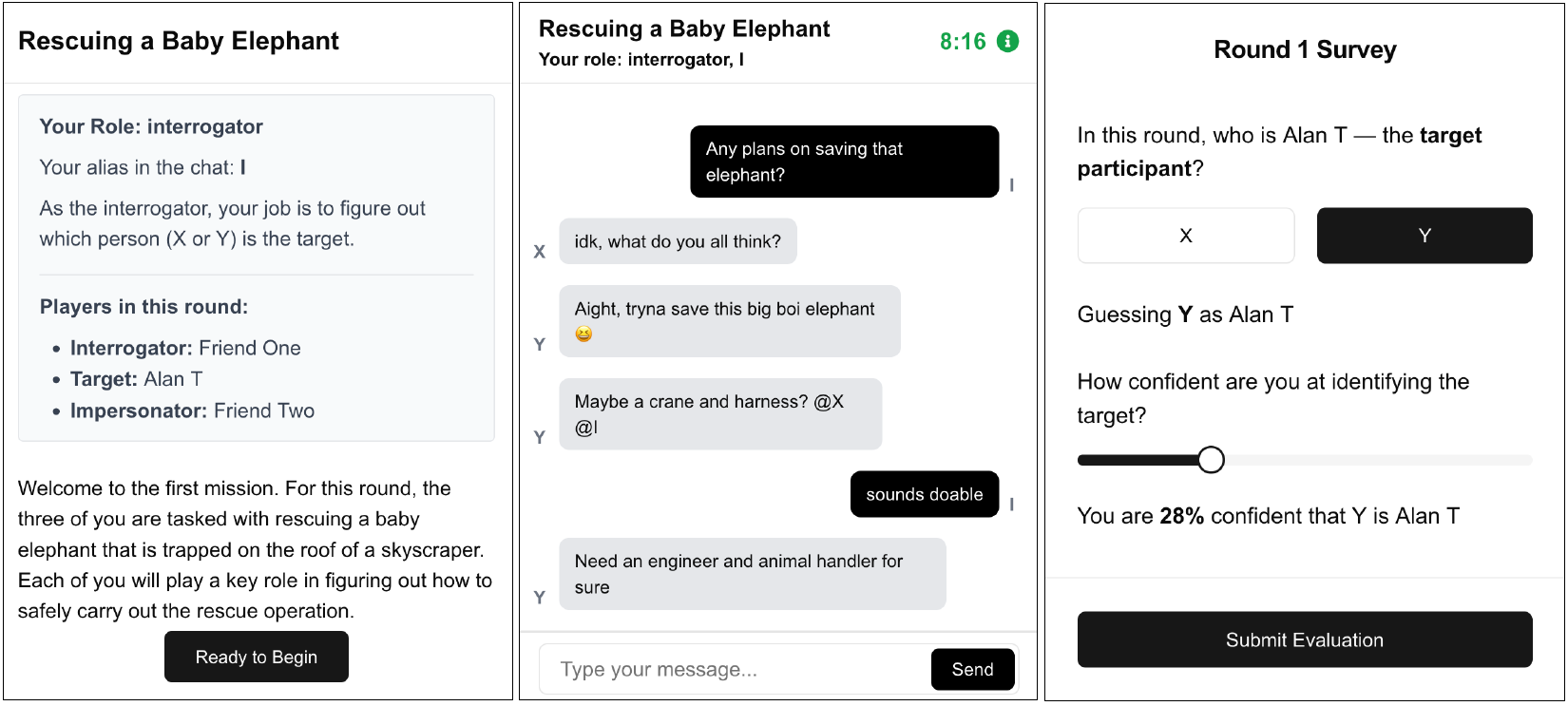}
	\caption{The chat interface powering the Personal Imitation Game. As the game begins, participants are presented with the first-round topic, complete player names, and their respective roles (left). During each game round, the interface automatically assigns random aliases for the impersonator and target, while the interrogator is shown as ``I'' (center). At the end of each round, the interrogator's guess and confidence level are recorded (right).}
	\label{fig:interface}
	\Description{Screenshot of a three-panel interface for the Personal Imitation Game.
		Panel 1: Round Details screen visible to all participants showing the round topic, the viewer's role and chat alias, a goal reminder, and a roster listing player display names with roles. A scenario prompt explains the mission. A Ready to Begin button advances to the chat.
		Panel 2: live chat during gameplay with a round title, role reminder, and countdown timer. Messages appear as labeled bubbles with sender labels (X, Y, or I).
		Panel 3: end-of-round survey for the interrogator, asking who is the target (X or Y) with a confidence slider.}
\end{figure*}

The Personal Imitation Game differs from both in protocol rather than in
subject. We restore Turing's three-party structure: the interrogator converses
with two players at once, the genuine target among them, and a human
impersonator supplies the competing baseline. Neither prior framework sets a
machine impersonator against a human impersonator of the same target, so neither
can ask the question we take as central --- whether a machine imitates a person
more convincingly than that person's own friends do. Both frameworks were
developed concurrently with, and independently of, the Personal Imitation Game. The same three-party structure distinguishes our work from \textit{personalized} agent frameworks~\cite{openclaw2025, openclawsite2025, liang2025pahf}, which learn user preferences to act on a user's behalf but do not attempt to replicate the user's conversational identity.

\subsection{Building Personal Chat Agents}

To build a digital clone, an LLM must be grounded in personal data, typically
via prompt engineering~\cite{language_brown_2020}, retrieval-augmented
generation (RAG)~\cite{lewis2020retrieval}, or fine-tuning~\cite
{howard2018universallanguagemodelfinetuning}, with domain specialization a
related strategy~\cite{beltagy2019scibertpretrainedlanguagemodel,
	ling2024domainspecializationkeymake}. Prior work on digital twins
relies primarily on prompt engineering~\cite
{park2024generativeagentsimulations1000,
	toubia2025twin2k500datasetbuildingdigital}: for instance, Park et al.\
prepended two hours of interview transcripts to an LLM context~\cite
{park2024generativeagentsimulations1000}; Toubia et al.\ included raw text
from a 500-question survey~\cite
{toubia2025twin2k500datasetbuildingdigital}. The prototype system presented
in this paper combines prompt engineering with a RAG system that queries a
user's repository of personal messages. Unlike systems that require
conducting new interviews or lengthy surveys~\cite
{park2024generativeagentsimulations1000,
	toubia2025twin2k500datasetbuildingdigital}, this approach leverages extant
messaging data, lowering the barrier to constructing personal agents and
improving the ecological validity of the grounding.

\section{The Personal Imitation Game}

\noindent To explore personal conversational agents, we designed a game
inspired by Turing's original Imitation Game~\cite
{computing_turing_1950}. Like the original test, each round $R$ of the game
involves three participants: an interrogator $i$, a target $t$, and an
impersonator $m$.
In our setup, the interrogator chats with both target and impersonator
in a shared private messaging room. The target participant is instructed to
communicate authentically, while the impersonator aims to mislead the
interrogator into believing that they are the target. At the end of the
round, the interrogator must guess who the real target is: if the
interrogator guesses correctly, they (and the target) win the game; if the
interrogator guesses incorrectly, the impersonator wins.
Each round of the game lasts until the interrogator is ready to make a guess, or ten minutes
have elapsed.

To foster active participation in the chat and prevent questions that trivialize the game (e.g., ``What did I give you for your birthday last year?''), the test provides a discussion topic for each round. 
We conducted two formative studies ($N=59$; Appendix~\ref{appendix:formative-studies}) to evaluate different types of tasks within the framework of the modified Turing test.
The results indicated that hypothetical collaborative planning tasks with universally understandable contexts (e.g., saving an animal) generated the richest discussions.
Based on these findings, we created three co-design scenarios that provide shared intent and context to provoke meaningful communication~\cite{searle_speech_acts, cohen_designing_adversarial_collab_2000}, while remaining general enough to avoid favoring any particular participant. 
In our second formative study, whose rounds allowed twenty minutes, interrogators consistently reached decisions within ten minutes, motivating the ten-minute upper bound on round duration (Appendix~\ref{appendix:formative-studies}).

Unlike Turing's original test --- which can be conducted with strangers --- the PIG 
requires relationships between participants: the
interrogator and impersonator must each have some personal familiarity with
the target in order to carry out their roles. To formalize this familiarity,
we draw from the social psychology literature on \textit{tie strength}~\cite
{granovetter_weakties}, whose best single indicator is closeness --- the
perceived emotional intensity of a relationship~\cite{marsden_measuring_1984}.
We treat tie strength as a dyadic property and measure it by averaging the two
participants' reciprocal estimates of that intensity.

To facilitate the game, we built a software platform (Figure~\ref{fig:interface}). It features a text-based chat interface for the
participants, automatically rotates and assigns roles and topics, anonymizes
participants with randomized aliases, and displays the time left in each
round. The software also allows researchers to observe the chats in
real time, collects the interrogator's guess and confidence estimate, and
shows the winners and losers of each game.

\section{Human Baseline Study} \label{study_section}

To establish a baseline for how people identify specific individuals in
real-time chats, we ran the PIG on ten groups of \textit{human} participants.
We report on their success at impersonation (and identifying impersonators),
the signals they use to make assessments, their confidence in doing so, and
how these variables change with the strength of the participants'
relationships.

\subsection{Procedure}\label{human_baseline_procedure}

To conduct the study, we briefed participants on its objectives and discussed data usage and privacy considerations. 
We explained the rules of the game, winning criteria, the topics to be discussed, and the compensation structure for participation.
The game was conducted either in person or via videoconference. All interaction was text-only and participants were isolated from one another, so we pooled the two modalities rather than testing for a difference between them.
After briefing, one researcher was assigned to each participant, and participants were guided to separate physical and virtual rooms accordingly (Figure~\ref{fig:pig}, left).
Each researcher remained available to answer questions and placed remote participants in individual Zoom breakout rooms with cameras and microphones disabled.
Prior to the start of the game, participants signed a consent form and made tie strength assessments for their group members.
All procedures received an IRB Exempt determination from our institution (Category 3; 45 CFR 46.104(d)(3)).

To incentivize active participation of all three group members in the game, we
awarded a \$10 monetary prize for each completed round. The target and
interrogator shared the goal of establishing the target's identity,
whereas the impersonator's objective was to deceive them. When the
interrogator guessed correctly, the interrogator and target were awarded \$5
each; when the interrogator guessed incorrectly, the impersonator was
awarded \$10. This incentive structure mirrors that of Turing's Imitation
Game~\cite{computing_turing_1950}. Participants were warned that attempts to
subvert the intent of the game (for instance, by asking personal questions
unrelated to the topic) would result in disqualification, forfeiting that round's prize; because the
incentive was purely performance-contingent, a participant who won no round
received nothing.

Participants were each given an ID and instructed to log in to our chat
interface. The interface randomly assigned roles to each player and chose one
of three pre-selected discussion topics for each round: \textit
{Elephant}: saving a baby elephant from the roof of a skyscraper; \textit
{Train}: stopping a runaway train to protect the people inside and in the
city; or \textit{Bomb}: safely defusing a bomb in a museum and evacuating the area~(Appendix~\ref{appendix:pig-topics}). 
Since each topic was tied to a fixed round, topic is confounded with round order. Our formative studies (Appendix~\ref{appendix:formative-studies}) established that all three scenarios reliably stimulate discussion, and we report aggregate performance over this fixed sequence, so topic and
round-order effects cannot be separated.
The interface then placed all three participants in
one chat room and assigned them aliases: ``I'' for the interrogator, while
the target and impersonator were randomly assigned ``X'' and ``Y'' to avoid any
confounds. When time expired --- or the interrogator was ready to make a
guess --- the interface provided a toggle button for selecting the target
participant (i.e., ``Who is the target, \textit{participant's name}: X or Y?'')
and a slider to indicate the confidence level. Roles were rotated clockwise, and
the process was repeated for three rounds; all participants experienced each
role once.

One researcher served as the facilitator for each game session, monitoring the
chat~ and enforcing the rules.
After each game, the facilitator conducted a verbal semi-structured group
interview with the participants to discuss the signals and strategies they
employed (Appendix~\ref{appendix:interview-guide}). The facilitator then debriefed the participants on the purpose of
the study and who won each round, then distributed rewards. We anonymized the interview transcripts and analyzed them using iterative
open coding, coding
inductively to discover signal classes, then applying a deductive pass
against Stamatatos's stylometric categories~\cite{stamatatos2009survey} and the
Big Five~\cite{John1999_big_five}. Four researchers built the codebook by
consensus, resolving disagreements in discussion rather than coding
independently, so no inter-rater agreement statistic is available and we treat the qualitative analysis as exploratory rather than as a reliability-checked measurement. 

The participant name--ID mapping is destroyed at study
close-out~\cite{tullis2021self}.

\subsection{Measures}

To understand the results of our study, we compute a number of useful
measures: the \textit{interrogator success rate} (ISR) for each group; the \textit
{interrogator confidence} for each game round, recorded on a 0--100 slider
in one-point increments; the \textit{tie strength} $T_{a\leftrightarrow b}$ of each
reciprocal interparticipant relationship $a\leftrightarrow b$ within a group;
and the \textit{group tie strength} $T_{G} = \operatorname{avg}_
{a \leftrightarrow b \,\in\, G} \left(T_{a \leftrightarrow b}\right)$ of all group
relationships. We report 95\% confidence intervals for all
interrogator success rates using the Wilson score interval~\cite{wilson1927probable}. These intervals assume independent rounds and do not account for
within-group dependence.
\noindent The confidence slider carried no endpoint labels, opened at 50, and
restated the selected value as the
probability that the chosen player was the target (Figure~\ref{fig:interface},
right).

To assess tie strength, each participant $a$ self-reported the perceived
emotional intensity $I_a(\cdot)$ of their relationship with each other group
member on a five-point Likert scale. Then, we define $T_
{a \leftrightarrow b} = \frac{1}{2} \left( I_a(a\leftrightarrow b) + I_b
(a\leftrightarrow b) \right)$ to be the average of the emotional intensities
reported by each participant in the relationship.

To explore the impact of relational dynamics on interrogator success, we also
compute the \textit{differential tie strength} $\Delta T_{R} = T_
{m \leftrightarrow t} - T_{i \leftrightarrow t}$ of each round $R$, which
captures the impersonator's advantage over the interrogator in
closeness to the target.

\subsection{Participants}

We recruited 10 groups --- each consisting of three
 participants --- snowball-sampled from the campus of a major US
 research university. 20 of the participants identified as men; 10 identified
 as women. Participant ages ranged from 19 to 34 (median = 21.5). 15
 participants were pursuing undergraduate degrees, 12 master's degrees, and
 three PhDs; 22 were Computer Science majors, while 8 were not. All
 participants in each group attested that they knew each other well, and were
 comfortable chatting in English. We label them P1--P30.
 Table~\ref{tab:baseline_participant_details} gives a per-participant breakdown
 (Appendix~\ref{appendix:baseline-demographics}).
 \subsection{Results} \label{human_baseline_results}

In 28 of the 30 rounds the interrogator used the full 10 minutes; in the
remaining two they guessed before time ran out.

Rounds averaged $64.4$ total
messages ($\sigma_{m} = 17.5$); messages averaged $7.0$ words in length
($\sigma_{w} = 2.0$). Engagement was similar across roles: targets averaged
$22.9$ messages per round ($\sigma = 9.5$); interrogators $22.0$ ($\sigma =
8.0$); and impersonators $19.6$ ($\sigma = 8.2$). Engagement was also
similar across tasks: the \textit{elephant} task averaged $64.3$ messages per
round ($\sigma = 20.6$); the \textit{train} task $65.7$ ($\sigma = 14.9$);
and the \textit{bomb} task $63.1$ ($\sigma = 18.4$).

\begin{figure}[th]
	\centering
	\includegraphics[width=\columnwidth]{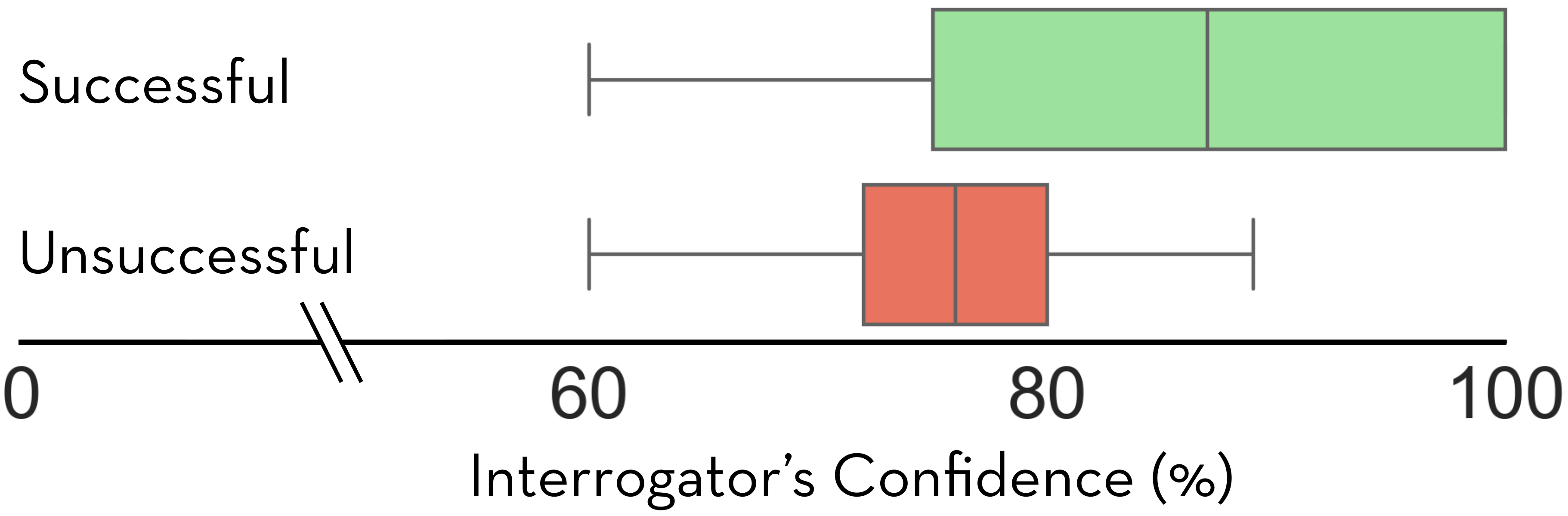}
	\caption{Successful interrogators have higher confidence (median = 87, IQR = [75, 100]) than unsuccessful interrogators (median = 76, IQR = [72, 80]). Whiskers are 1.5$\times$IQR fences and the horizontal axis is broken between 0 and 60. Three unsuccessful-interrogator values fall outside the fences and are not plotted (19, 50, and 100); they are included in the means and standard deviations reported in the text, which is why the reported $\sigma = 19.9$ is larger than the plotted spread suggests.}
	\label{fig:confidence}
	\Description{Boxplot comparing interrogators' confidence (\%) by outcome. Two horizontal boxplots are shown side-by-side: the top boxplot represents Successful interrogators (green) and the bottom boxplot represents Unsuccessful interrogators (salmon). Successful: median = 87\%, interquartile range (IQR) = 75–100\%. Unsuccessful: median = 76\%, IQR = 72–80\%. Medians are marked by a line inside each box; whiskers extend to the furthest value within 1.5 times the IQR, so the unsuccessful whiskers run from 60 to 89 and three values outside those fences (19, 50, 100) are not drawn. The horizontal axis at the bottom is labeled "Interrogator's Confidence (\%)" and is broken: it shows 0, then jumps to the 60--100 range in which the boxes lie. Category labels ("Successful," "Unsuccessful") appear on the vertical axis so the figure does not rely on color alone.}
\end{figure}

\medskip
\noindent\textbf{Success Rate \& Confidence}\\
\noindent Out of the 30 games played, interrogators successfully identified
the target in 17 instances, resulting in a success rate of 56.7\% (95\% Wilson CI $[0.39, 0.73]$). This rate is not distinguishable from chance (exact binomial $p = 0.58$); we treat it as a descriptive reference point rather than as evidence that interrogators can identify a target better than a coin flip. Since target and interrogator success is reciprocal, this is
 also the rate at which targets were able to prove their identity. In the
 remaining 13 games --- or 43.3\% --- impersonators succeeded in mimicking
 the target. Of the 30 participants, four earned \$20, six earned \$15, nine
 earned \$10, eight earned \$5, and three lost all the rounds they played.

Successful interrogators reported an average confidence level of 85.9\%
($\sigma = 13.5\%$), with a median of 87\% (IQR = [75, 100]), indicating a high degree of certainty in their
judgments. In contrast, unsuccessful interrogators had an average confidence
level of 71.9\% ($\sigma = 19.9\%$), with a median of 76\% (IQR = [72, 80]). This 14 percentage-point difference (two-sample $t$-test, $t = 2.31$, $\mathit{df} = 28$, $p = 0.028$)
suggests an association between reported confidence and accuracy in this
sample (Figure~\ref{fig:confidence}). This association does not establish that
confidence reports were calibrated probabilities. Confidence is bounded above
and 5 of the 17 successful interrogators sat at the 100\% ceiling, so this test
should be read as approximate. It also assumes independent observations and
does not account for within-group dependence.

\begin{figure}[tb]
	\centering
	\includegraphics[width=\columnwidth]{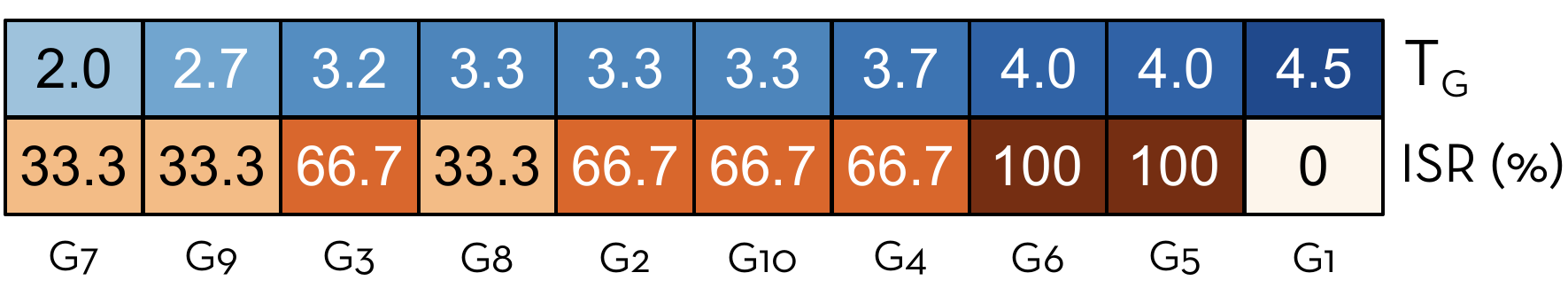}
	\caption{A heatmap showing the relationship between group tie strength $T_{G}$ (top-blue) and interrogator success rate (bottom-orange). Groups are sorted left to right from lowest to highest $T_{G}$.}
	\label{fig:group_tie_strength}
	\Description{Heatmap with two rows comparing group tie strength and interrogator success rate across ten groups, ordered left-to-right by increasing tie strength. Columns (left to right): G7, G9, G3, G8, G2, G10, G4, G6, G5, G1.
		Top row (blue) encodes group tie strength $T_G$ with values: G7=2.0, G9=2.7, G3=3.2, G8=3.3, G2=3.3, G10=3.3, G4=3.7, G6=4.0, G5=4.0, G1=4.5.
		Bottom row (orange) encodes interrogator success rate (\%) with values: G7=33.3, G9=33.3, G3=66.7, G8=33.3, G2=66.7, G10=66.7, G4=66.7, G6=100, G5=100, G1=0.
		The x-axis labels groups by ID; the two-row layout places $T_G$ directly above the corresponding success rate for each group. The figure shows that higher tie strength at or above 4.0 is associated with extreme outcomes---either 100\% success (G6, G5) or 0\% (G1).}
\end{figure}

\medskip
\noindent\textbf{Tie Strength}

 \noindent Across all groups, the average $T_{G}$ was 3.4
 ($\sigma = 0.71$). Among the ten groups, the three with the highest tie
 strength were also the three with the
 most extreme interrogator success rates (Figure~\ref{fig:group_tie_strength}): Group 1 ($T_{G} = 4.5$) was the only group with a
 0\% interrogator success rate, while Group 5 ($T_{G} = 4.0$) and Group 6 ($T_
 {G} = 4.0$) were the only groups with a 100\% interrogator success rate.
 With ten groups we read this as a pattern in these data rather than a
 relationship we have tested.
 Close relationships within a group can make it easier for an interrogator to
 differentiate between the target and impersonator, but also make it easier
 for an impersonator to mimic the target.

These data do not settle whether differential tie strength predicts
 interrogator success. Across all thirty rounds the point-biserial correlation
 between $\Delta T_{R}$ and interrogator success is $r = -0.27$ (95\% CI
 $[-0.53, 0.08]$; a permutation test respecting the nesting of rounds within groups
 gives $p = 0.23$), and the three rounds taken separately disagree in sign. At thirty rounds, a two-tailed test at $\alpha = .05$ has 80\% power only for
$|r| \geq 0.49$, so we report the estimate and draw no conclusion from it.

\begin{figure}[tbp]
	\centering
	\makebox[\columnwidth][l]{%
\includegraphics[width=1.08\columnwidth]{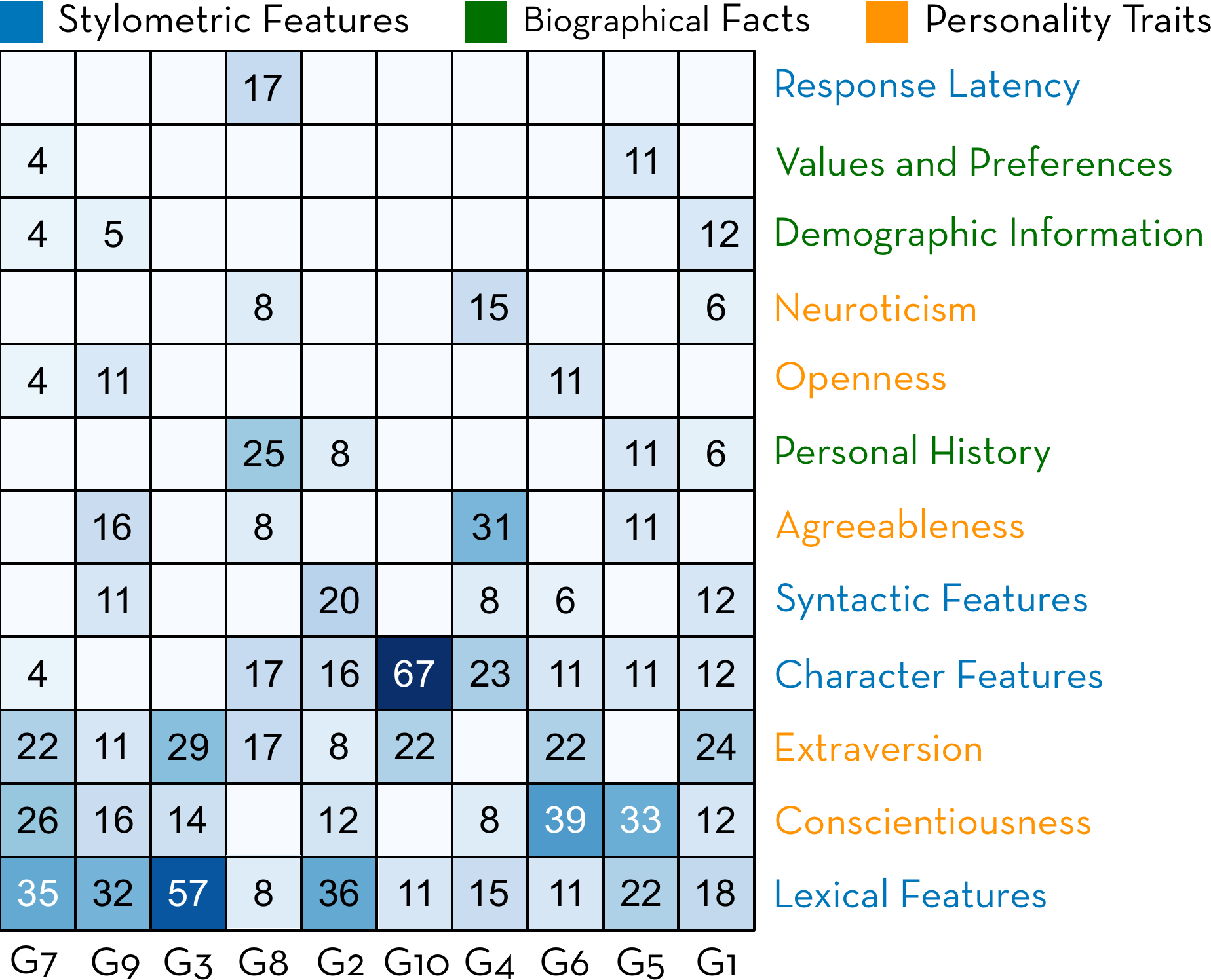}%
}	\caption{The distribution of signals used by participants across groups, sorted left to right from lowest to highest $T_{G}$. Each cell gives that signal's share of the group's coded signal mentions, so columns sum to 100\%; darker shades indicate higher shares. Because per-group totals differ (7 to 25 mentions), shares are not comparable across columns as raw counts.}
	\label{fig:code_group_heatmap}
	\Description{Heatmap of signal usage across ten groups (G1--G10), with columns ordered left-to-right by increasing tie strength: G7, G9, G3, G8, G2, G10, G4, G6, G5, G1. Rows list 12 identity signals (ordered top to bottom by increasing frequency in the data). Row labels are color-keyed by category---stylometric features (blue), personality traits (orange), biographical facts (green)---while the heatmap cells use a single intensity scale (light to dark) for normalized frequency. Row labels (top to bottom as shown): Response Latency; Values and Preferences; Demographic Information; Neuroticism; Openness; Personal History; Agreeableness; Syntactic Features; Character Features; Extraversion; Conscientiousness; Lexical Features. Notable patterns: G3 concentrates on Lexical Features (57\% of that group's coded mentions); G10 concentrates on Character Features (67\%). Signal breadth is unrelated to tie strength: G1 uses the most distinct signals (8 of 12) and G7, the lowest tie-strength group, uses 7. Category colors appear on row labels so interpretation does not rely on cell color alone.}
	\vspace{-.03in}
\end{figure}

\smallskip
\noindent\textbf{Identity Signals}\\
\noindent After conducting an iterative open coding of the post-game interview
 transcripts, we found that participants predominantly employ three classes
 of signals to identify individuals in text-based conversation (Figure~\ref{fig:code_group_heatmap}). For each of
 these classes, we created a more granular codebook by leveraging established
 categories already present in the literature.

\textit{Stylometric features}, comprising distinguishing linguistic traits
 such as vocabulary, tone, and grammatical structure, were mentioned
 frequently: ``I'd use punctuation and capitalization'' (P4); ``mine is a
 little more formal with no shorthands'' (P12). Our codebook follows the stylometric framework by Stamatatos~\cite{stamatatos2009survey}:
\begin{itemize}[leftmargin=*]
  \item \textit{Lexical features:} refer to vocabulary and word choice, including sentence length and word/phrase frequency
  \item \textit{Syntactic features:} focus on sentence structure, part-of-speech tag sequences, and tense consistency
  \item \textit{Character features:} focus on character sequencing such as typos, unique spellings, and emojis
  \item \textit{Response latency:} refers to texting speed and response time
\end{itemize}

\textit{Personality traits} also provide useful identity signals that may be
 gleaned from text, for instance, by espousing empathy or a predisposition
 towards humor: ``so I tried showcasing sympathy towards the elephant''
 (P10); ``he'll try to have a funny reply to everything'' (P3). Our
 codebook leverages the Big Five Personality Traits framework from
 psychometrics~\cite{John1999_big_five}:
\begin{itemize}[leftmargin=*]
  \item \textit{Openness:} signs of curiosity, spontaneity, and inquisitiveness
  \item \textit{Conscientiousness:} indications of diligence, organization, and responsibility 
  \item \textit{Extraversion:} reflections of sociability 
  \item \textit{Agreeableness:} expressions of positive emotions, empathy, and caring attitudes
  \item \textit{Neuroticism:} manifestations of emotional instability or sensitivity
\end{itemize}

\textit{Biographical facts} are also used to imitate and identify others in the
 game: ``so I'm a biology major, and I just kind of used some of the words
 that I've used before and that relate to my major, which gave an indication
 of who I am'' (P6); ``yeah, they just know many things in general about
 football'' (P19). Our codebook encodes:
\begin{itemize}[leftmargin=*]
  \item \textit{Demographic information:} age, gender, cultural background, and other demographic details
  \item \textit{Personal history:} references to significant life experiences, education, or social connections
  \item \textit{Values and preferences:} statements reflecting values, ethical principles, beliefs, and personal preferences
\end{itemize}

Leveraging the codebook, we examined the distribution of identity signals
across groups. We found that most groups used four signals --- lexical
features, conscientiousness, extraversion, and character features --- to
reason about identity during the game. Other signals, like values and preferences,
neuroticism, response latency, and demographic information were rarely
utilized. Signal breadth is not related to tie strength: G1, the group with
the highest $T_{G}$, drew on the widest range of signals (8 of 12), but G7 --- the
group with the \textit{lowest} $T_{G}$ --- used 7, as did G8 and G9, more than
either G5 or G6 (6 each). The two narrowest groups, G3 and G10 (3 signals each),
sit mid-range on $T_{G}$. Across the ten groups the rank correlation between
breadth and tie strength is negligible ($\rho \approx 0.05$). We report which
signals interrogators invoked and how often, and do not relate individual
signals to whether the interrogator guessed correctly.
\smallskip

\section{Building an AI Impersonator} \label{system_section}

Having established a human baseline for performance in the PIG, we next ask
the natural question: can we build an LLM-based system that can impersonate a
specific individual in real-time chat? To construct such a system, we prompt
a foundational language model with the three signals identified in our study:
stylometric features, personality traits, and biographical facts. To ground
these signals, the paper's lead author conducted an autoethnography yielding
four outputs: an archive of historical messages, a Big Five personality
assessment, a stylometric self-assessment, and an autobiography.
He is also the person the system impersonates: we call him the
\textit{target author} when discussing the game, and the \textit{lead author}
when discussing the research. This single-subject approach is a deliberate methodological choice~\cite
{lucero2019sampleofone}: a personal agent demands longitudinal data that is
impractical to collect from third parties, making the researcher's own
digital history the natural starting point for establishing feasibility.

To play the PIG, we deploy a RAG system that first selects relevant
 historical messages from the author's archive based on the current chat
 state. A prompt construction module then combines these retrieved messages
 with the chat, personality profile, stylometric profile, and autobiography
 to produce a single prompt instructing the model to evaluate and respond to
 the conversation. In the following subsections, we describe the
 autoethnographic methodology and dataset, the RAG setup and prompt design,
 the real-time chat implementation, and the safeguards used to handle
 third-party data.

\subsection{Autoethnography}\label{autoethnography_section}

To build a grounded model of individual identity, we adopt an autoethnographic
approach~\cite
{cunningham-autoethnograpy-05,ellis-authoethnography-10}. Autoethnography is
a qualitative research method in which the researcher critically reflects on
personal experiences to generate empirical insights~\cite
{jain-autoethnography-19}. Recent work in HCI applies it to first-person accounts of living with generative AI~\cite{moore-autoethnography-25}.
In this study, the paper's lead author --- an ethnically Indonesian male, 29 years old,
who has lived in the United States for 12 years at the time of writing --- served as the researcher and subject.
To support the design and evaluation of the AI Impersonator, the author compiled
four forms of self-representative data:

\begin{itemize}[leftmargin=*]
	\item \textit{Historical messages}: A corpus of personal digital communication spanning different social contexts and six messaging platforms
	\item \textit{Personality scores}: Results from a self-assessment based on the Five-Factor Model~\cite{johnson2014ipipneo120}
	\item \textit{Stylometric profile}: A self-assessment highlighting distinctive linguistic features, based on established stylometric frameworks~\cite{stamatatos2009survey}
	\item \textit{Autobiography}: A narrative describing biographical details, values, and formative life experiences
\end{itemize}

These four data streams operationalize the three identity signals
 uncovered in our human baseline study (Section~\ref
 {human_baseline_results}). Stylometric features come from the historical
 messages and the stylometric profile: the archive shows actual word choice,
 syntax, and pacing in context, while the profile summarizes the most
 distinctive patterns. Personality traits are captured by the Big Five
 personality scores, which are converted into short verbal descriptions and
 supported by examples from the messages and autobiography. Biographical
 facts come primarily from the autobiography, which lists demographic
 details, roles, and key life events, and is supplemented by contextual
 details in the historical messages.

This structured dataset serves as the foundation for building and evaluating
the AI Impersonator. This approach draws conceptual inspiration from
historical diaries, such as those of Samuel Pepys~\cite{pepys_1896}, framing
the long-form personal archive as a medium for building a computational model
of identity.

The historical message archive is used in two ways: to model how the author
communicates in written conversation, and to capture the \textit
{context} --- shaped by surrounding messages from \textit
{relational others}~\cite{lapadat_ethics_autoethnography} --- for the
author's communication. Since this context consists of third-party data
which may contain personally identifiable information (PII) or other private
content, it must be handled with extreme care. 

Accordingly, we follow the autoethnographic ethical guidelines from Lapadat,
Chang, and Tullis~\cite
{lapadat_ethics_autoethnography, chang2016autoethnography, tullis2021self}.
No third-party data is viewed or accessed by any researcher other than the
lead author, and raw messages never leave the author's personal device;
messages and their metadata are de-identified before use, and only the
de-identified, preprocessed messages reach the author's private Gemini
developer account and the Vertex AI index;
sensitive conversations are manually filtered from the dataset during
preprocessing; no attempt is made to link, learn, generalize, correlate, or
interrogate any personal information about the third-party conversants; and
human-in-the-loop moderation is used as a backstop to reduce the risk that private
data is inadvertently ``leaked'' during the PIG.

\medskip
\noindent\textbf{Historical Messages}\\
\noindent Messages were collected from six platforms, comprising 4,637
 distinct conversations and 434,085 messages between 2013 and 2024;
 Table~\ref{tab:historical_record_stats} gives the platform composition. Together, these platforms span a broad range of social contexts --- friends and coworkers (iMessage, Messenger, Instagram, WhatsApp), family (iMessage, WhatsApp), and romantic interests (Instagram, Tinder, Hinge\footnote{Dating platforms only export the author's sent messages, not the other party's.}).
Conversations divide almost evenly by span, 2,297
 beginning and ending within a single day against 2,340 running longer. The
 archive is predominantly English, the author sent 38.9\% of its messages, and
 it covers 2,949 unique conversants (individuals with multiple accounts across
 different platforms were disambiguated by hand).
 Table~\ref{tab:conversation_stats} gives per-conversation summary statistics.

\begin{table}[t]
 	\centering
 	\setlength{\tabcolsep}{3pt}
 	\begin{tabular}{@{} l r r @{}}
 		\textbf{Platform} & \textbf{Conversations} & \textbf{Messages} \\
 		\midrule
 		Facebook Messenger & 30.58\% & 64.51\% \\
 		Apple iMessage     & 28.25\% & 17.82\% \\
 		Tinder             & 26.57\% &  1.20\% \\
 		Instagram          & 13.05\% & 11.21\% \\
 		Hinge              &  1.10\% &  0.06\% \\
 		WhatsApp           &  0.45\% &  5.19\% \\
 	\end{tabular}
 	\caption{\label{tab:historical_record_stats}Platform composition of the historical chat record. Shares are of the 4,637 conversations and of the 434,085 messages.}
 	\Description{Table of platform composition for the historical chat record, with columns: share of conversations, share of messages.
 		Facebook Messenger 30.58 percent and 64.51 percent; Apple iMessage 28.25 and 17.82; Tinder 26.57 and 1.20; Instagram 13.05 and 11.21; Hinge 1.10 and 0.06; WhatsApp 0.45 and 5.19.}
 \end{table}

\begin{table}[t]
 	\centering
 	\vspace{-14pt}  %
 	\setlength{\tabcolsep}{3pt}
 	\begin{tabular}{@{} l r r r r r r @{}}
 		\textbf{Metric} & \textbf{Mean} & \textbf{SD} & \textbf{Q1} & \textbf{Median} & \textbf{Q3} & \textbf{Max} \\
 		\midrule
 		Duration (\textit{days})    & 227.7  & 502.7   & 0.4 & 5.1 & 159.1  & 3,454 \\
 		Messages (\textit{count}) & 93.6   & 504.4   & 2.0 & 5.0  & 20.0  & 10,004 \\
 	\end{tabular}
 	\caption{\label{tab:conversation_stats}Summary statistics per conversation for the historical chat record. We compute duration over the $N = 3{,}485$ conversations containing at least two messages and a nonzero span, and message counts over all 4,637.}
 	\Description{Two-row table of summary statistics from the historical chat record, with columns: Mean, Standard Deviation, Q1, Median, Q3, Maximum.
 		Row 1 --- Duration (days): mean 227.7; SD 502.7; Q1 0.4; median 5.1; Q3 159.1; max 3,454.
 		Row 2 --- Messages (count): mean 93.6; SD 504.4; Q1 2.0; median 5.0; Q3 20.0; max 10,004.
 		The two rows use different denominators: duration over the 3,485 conversations with at least two messages and a nonzero span, message counts over all 4,637.}
	\vspace{-0.4in}
 \end{table}

We obtain Apple iMessages using \texttt{imessage\-exporter}~\cite
{imessage_exporter}, which accesses the local SQLite iMessage database on the
author's Mac. WhatsApp conversations are exported directly from the app
using the ``Export Chat'' feature~\cite{wa_export_chat_faq}. Due to data
privacy regulations like GDPR~\cite{gdpr}, social platforms must
provide users with access to download their personal data,
including message records. We requested Instagram and Facebook Messenger logs
via Meta's ``Download Your Information'' portal~\cite
{meta_download_information}, and retrieved Tinder and Hinge data through
their respective account settings~\cite
{tinder_request_data, hinge_request_data}.

After collection, we post-process all message data by normalizing timestamps,
sender names, and message content into a canonical format. We remove links,
multimedia content, sensitive content, and automated system messages
(e.g., notifications from delivery services). Each sender name is anonymized
by mapping it to a unique identifier, and the mapping to original names or
phone numbers is deleted. Processed messages are permanently destroyed at
study close-out.

\medskip

\noindent\textbf{Personality Self-Assessment}

\noindent To generate the personality profile used by the AI Impersonator, the
 author completed the IPIP-NEO-120~\cite{johnson2014ipipneo120}, a
 public-domain Five-Factor inventory. The test, which took approximately 20 minutes to
 complete, includes 120 multiple-choice questions that measure five
 high-level traits and their 30 associated facets (six per trait).

Four items score each facet on a five-point scale, so facet scores run
4--20 and the six facets of a trait sum to 24--120. The prompt in
Appendix~\ref{appendix:prompt} presents these as 0--20 and 0--120; we report it
as deployed, and note that the lower end of each stated range is unattainable. Following the method of Serapio-García et al.~\cite
{personality_safdari_2023}, we converted these scores into natural language
descriptions using Goldberg's bipolar adjective markers~\cite
{goldberg_1992_devmarkersforbigfivefactor} (see Appendix~\ref{appendix:personality-binning} for the full binning scheme).
This mapping produces an interpretable personality summary that can be
embedded in the LLM prompt (Appendix \ref
{appendix:prompt}).

\medskip

\noindent \textbf{Stylometric Self-Assessment}\\
\noindent To capture the author's linguistic fingerprint, we composed a
 stylometric profile based on three commonly used classes of features in
 authorship analysis: lexical, syntactic, and character-level~\cite
 {stamatatos2009survey, jin_deep_2021}. These stylometric observations form part of the input
 signal for the AI Impersonator, enabling it to better reproduce the author's
 individual writing style.

\begin{figure*}[tbp]
	\centering
	\includegraphics[width=\textwidth]{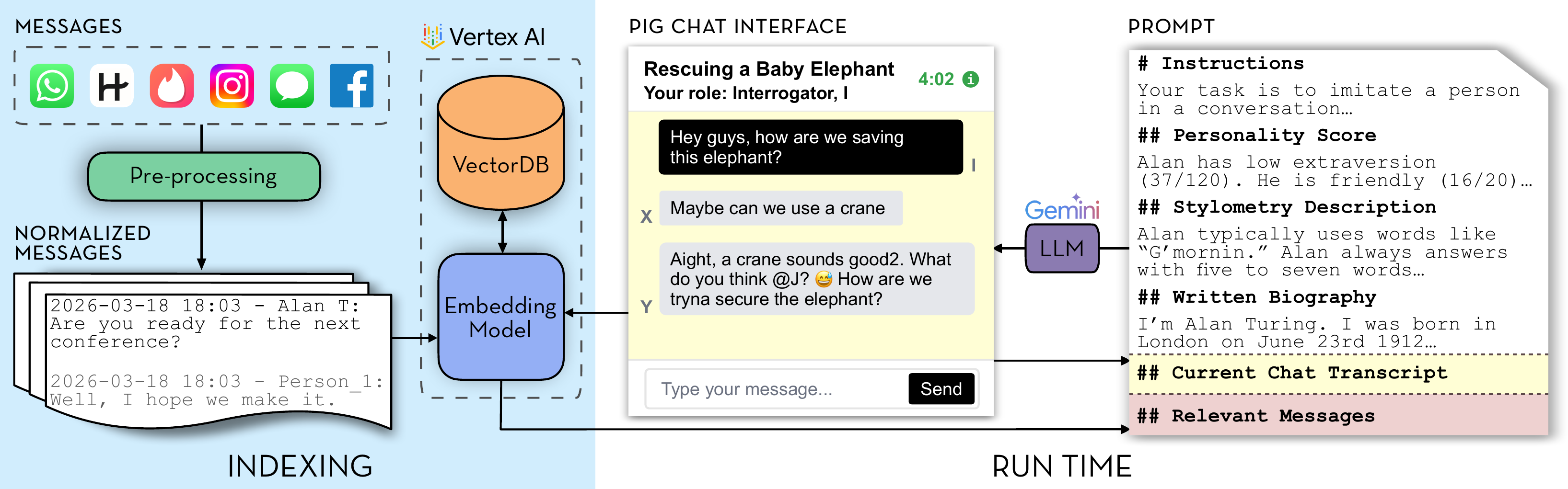}
	\vspace{-.1in}
	\caption{The system architecture of the AI Impersonator. Historical messages are preprocessed and indexed for retrieval in a vector database (left). A dynamically constructed prompt combines static identity signals with relevant messages from the history and the current PIG chat context (right-top). The LLM updates the chat in real time (right-bottom).}
	\Description{System architecture diagram in three panels (left, center, right) illustrating the AI Impersonator pipeline from data ingestion to live chat response.
		Left panel (Data ingestion and normalization): personal messages are pulled from multiple platforms (WhatsApp, Hinge, Tinder, Instagram, iMessage, Facebook). An arrow leads into a Pre-processing node. The output is a stack of Normalized Messages, each record containing a timestamp, sender name, and message text. Third-party senders are anonymized (e.g., Person\_1). An arrow leads from Normalized Messages to the Vertex AI platform, where messages are embedded by an embedding model and stored in a vector database for retrieval.
		Center panel (Prompt assembly for the AI Impersonator): the prompt includes (1) instructions to imitate a specific person in conversation; (2) three identity signals for that person: Personality scores, Stylometric profile, and a short written autobiography; (3) the current PIG chat transcript; and (4) relevant historical messages retrieved from Vertex AI via vector search. This assembled prompt is sent to the Gemini LLM.
		Right panel (Game live chat interface and response loop): the chat UI displays the round title, a role reminder, and a countdown timer. Messages appear as labeled bubbles. When a new user message is sent, the AI Impersonator is triggered; Gemini generates a reply that is returned to the chat, completing the loop.}
	\label{fig:system-architecture}
	\vspace{-.1in}
\end{figure*}

\medskip

\noindent \textbf{Autobiography}\\
\noindent Autobiographical self-description plays a key role in personal
 identity modeling. Inspired by historical first-person narratives like Saint
 Augustine's \textit{Confessions}~\cite{augustine1838confessions} and the
 diaries of Samuel Pepys~\cite{pepys_1896} --- and supported by recent work
 showing that personal reflection can guide LLM behavior~\cite
 {park2024generativeagentsimulations1000} --- we asked the author to write an
 extended account of his own life. 

The author composed a 2,400-word autobiography over the course of an hour. His
narrative includes demographic attributes (e.g., age, gender, cultural
background), personal history (e.g., education, social connections, formative
experiences), and value systems (e.g., ethical beliefs, preferences,
religious and political orientation). This data provides biographical context
and factual grounding for the AI Impersonator.

\subsection{System Design \& Implementation}

The AI Impersonator is built on a RAG framework, using an LLM guided by a
carefully constructed prompt. It integrates historical messages, personality
traits, the stylometric profile, and biographical details to generate
responses in real time to play the Personal Imitation Game.

The system consists of three components (Figure~\ref
{fig:system-architecture}): (1) a RAG module that indexes and retrieves
relevant historical messages based on the current PIG chat state, (2) a
prompt builder that combines the static personality, stylometry, and
biographical autoethnography data with the historical context from the RAG
and the current PIG chat, and (3) an LLM that generates a real-time response
using the provided prompt.

\medskip

\noindent\textbf{Retrieval-Augmented Generation}\\
\noindent To enable RAG, each de-identified conversation extracted from the historical
 message data is uploaded to Vertex AI~\cite{vertex_ai}, and segmented into
 500-token chunks for indexing into a vector database. During inference, the
 system uses the current PIG chat state as a query against this database, and
 retrieves the 50 most semantically similar messaging chunks, which are
 folded into the prompt to provide contextual grounding. Indexing the full
 messaging dataset took approximately four hours; at inference, RAG retrieval
 latency averages 1.1 seconds per query. The paid Gemini API
terms state that Google does not use prompts or responses to improve its
products~\cite{google_gemini_terms_2025}. We note the limits on reproducing this
system. We chose the chunk size of 500 tokens and the retrieval depth of $k = 50$
without tuning, and ran no sensitivity analysis over either. The
processed corpus and embedding index were destroyed at study close-out and the
model checkpoint is deprecated, so the deployment cannot be re-run by anyone,
including ourselves; the prompt template in Appendix~\ref{appendix:prompt} is
the reproducible artifact. 

\smallskip

\noindent\textbf{Prompt Construction}\\
\noindent The LLM prompt (shown in Appendix~\ref{appendix:prompt}) comprises
 two components: a static portion, and a dynamic portion. The static portion
 gives the LLM instructions on its task --- detailing PIG chat context and
 instructing the model to stay on topic and avoid hallucinating unverifiable
 facts or biographical details --- and explicitly disables
 ``AI assistant''-like behavior by forbidding the use of markdown, bullet
 points, or repeated messages. It also provides explicit guidelines
 for \textit{how} to impersonate the target --- using the three signals
 identified in our human baseline study --- and includes the personality
 scores, stylometric profile, and autobiography. The prompt is goal-free
 with respect to the game: it never mentions an interrogator, a rival
 impersonator, deception, or a win condition, and it supplies a
 \texttt{<Time Left>} value without explaining why time matters. Concretely, the coded
identity signals from our human baseline are operationalized as prompt
content: each personality facet is rendered as a natural-language descriptor
via the binning scheme in Appendix~\ref{appendix:personality-binning}
(following Serapio-Garc\'ia et al.~\cite{personality_safdari_2023}), so that
signals such as extraversion and conscientiousness are steered explicitly,
while the stylometric profile and autobiography are inserted directly. The dynamic portion is
 assembled at runtime after each new message is received in the PIG,
 comprising the current PIG state (i.e., transcript, topic,
 alias, time left) and the 50 historical message segments retrieved via the
 RAG.

\smallskip
\noindent\textbf{Implementation}\\
\noindent To simulate natural human typing, we introduced a stochastic delay
 mechanism in the chat system. After receiving a message, the system waits
 ten seconds, or the model's generation time if that is longer, plus a small
 random offset, before initiating a response (simulating reading and thinking). After a reply is
 generated, characters are rendered at a 40 WPM typing speed. Messages from
 other participants received while the model is generating are held until the
 response is complete, to ensure temporal coherence and prevent overlap
 during rapid turn-taking.

We used Gemini 1.5 Pro~\cite{reid2024gemini15} as the system's LLM due to its large context
window, competitive benchmark performance~\cite
{mmlupro_wang_2024}, and low latency. At the time of deployment, Gemini
supported up to 1 million tokens of context~\cite{reid2024gemini15}. Each PIG prompt contained about
31k tokens: roughly 1,500 in the instructions, 130 in the game
 state, 400 in the personality scores, 600 in the stylometric profile, 3,300
 in the autobiography, and 25k in historical messages. The cost per
 Gemini API call was 3.9\textcent. On average, reply generation
 took about 11 seconds, with a total interaction latency of $\sim$25 seconds (including RAG latency and simulated typing delay).

\section{AI Impersonator Study}

Having built an AI Impersonator, we use it to play the PIG in a second,
human-machine study. The study seeks to answer two related research
questions. First, can the AI Impersonator imitate the target author more
convincingly than a human impersonator who has a strong interpersonal
relationship with him? Second, can the AI Impersonator convincingly imitate
the target author when placed in a head-to-head test against him?

\subsection{Procedure}

To address these questions, we adapted the PIG into a deception study under a separate IRB protocol, which authorized withholding from participants that one player was an AI until the post-session debrief. Each
game involved two participant friends of the target author, playing three
rounds of PIG. In Round 1, the AI Impersonator replaced the target author,
one participant was the impersonator, and the other the interrogator
(Figure~\ref{fig:pig}, right), with the two roles
assigned by recruitment order. 
In Round 2, we conducted a standard
 --- throwaway --- PIG round to preserve the deception, with one participant
   randomly assigned to be the target (Figure~\ref{fig:pig}, left). Round 3
   consisted of two parallel games, with both participants being told
   (deceptively) they were playing together. In truth, each was paired
   independently with the target author and his AI Impersonator (Figure \ref
   {fig:pig}, center). The two games ran concurrently: the target author
   held both chats open side by side and answered each in turn, while a separate
   instance of the AI Impersonator served each game. Participants were not told
   that the round was structured this way. The author received no instruction
   beyond playing as himself. This design allowed for a controlled evaluation of
   both research questions, without disclosing the presence of an AI to the
   participants.

Each AI Impersonator session was live-moderated by a facilitator, who could
halt the game if participants began attempting to discuss PII, or if the AI's
output ``leaked'' any third-party data (for instance, by inadvertently
mentioning a friend's name). No such incidents occurred during the study.
After the games were completed, researchers interviewed the participants as
in the human baseline study (Appendix~\ref{appendix:interview-guide}). The
guide additionally included a closing question on AI suspicion (question~4),
asked before the debrief to measure how many participants detected the
deception. Researchers then conducted a scripted debrief disclosing the
AI involvement.

\vspace{-.05in}

\subsection{Participants}

We recruited 10 groups of 2 participants, all mutual friends of
 the target author. All participants were men between 24 and 41 years of
 age (median = 28.5) and fluent in English, and none had prior exposure to the PIG
 study or knowledge of the AI Impersonator. They had known the target author for
 a mean of 6.45 years (range 2--13 years), with a mean group tie strength of
 3.62 against 3.40 in the human baseline. Numbering continues from the baseline
 study: we label these participants P31--P50, and the two cohorts are disjoint. Seven participants were software
 engineers; six were PhD students; one was a master's student in Computer Science; three
 were startup founders; and three held other roles
 (analyst, consultant, sales).
A detailed breakdown of participant demographics is provided in Table~\ref{tab:participant_details} (Appendix~\ref{appendix:participant-demographics}).
\subsection{Results} \label{ai-study:results}

\noindent The design yields 40 games across 30 group-rounds: ten in
Round 1, ten in Round 2, and twenty in Round 3, whose two games run in parallel
per group. Of these, 34 ran for the full ten minutes. The remaining six ended
early --- three in Round 1, one in Round 2, and two in Round 3 --- with a mean of
4 minutes and 45 seconds still on the clock ($\sigma$ = 1 minute 29 seconds).
Each game produced an average of 43.4 messages ($\sigma = 17.0$), with a mean
message length of 7.9 words
($\sigma = 8.2$). 
The target author sent an average of 11.8 messages in Round 3 ($\sigma = 5.2$), the AI Impersonator 14.0 in Round 1 ($\sigma = 5.5$), the human impersonator 10.3 in Round 1 ($\sigma = 5.6$), and the interrogators 13.5 across all rounds ($\sigma = 8.4$).
Message volume also varied by task, with the \textit{elephant} task averaging $36.5$ messages per game ($\sigma = 17.1$), the \textit{train} task $52.3$ ($\sigma = 14.5$), and the \textit{bomb} task $42.3$ ($\sigma = 17.0$).

\medskip

\noindent\textbf{Success Rate \& Confidence}\\
\noindent The interrogator success rates across all three PIG variants
 (including the human baseline) are summarized in Table \ref
 {tab:success_rate_comparison}.

 \begin{table}[tbp]
	\centering
	\setlength{\tabcolsep}{3pt}
	\begin{tabular}{@{} l c c r r r @{}}
		\textbf{Rounds} ($N$) & \textbf{Target} & \textbf{Impersonator} & \textbf{ISR} & \textbf{AI win} & \textbf{Conf.} \\
		\midrule
		\multicolumn{6}{l}{\textit{Human Baseline}} \\
		\quad All (30) & Human & Human & 56.7\% & --- & 79.8\% \\
		\addlinespace
		\multicolumn{6}{l}{\textit{AI Study}} \\
		\quad 1 (10) & AI & Human & 70.0\% & 70.0\% & 75.0\% \\
		\quad 2 (10) & Human & Human & 50.0\% & --- & 50.0\%\rlap{$^{\dagger}$} \\
		\quad 3 (20) & Author & AI & 65.0\% & 35.0\% & 76.7\% \\
	\end{tabular}
	\caption{\label{tab:success_rate_comparison}
		Interrogator performance across the four PIG rounds. In the baseline and Round 3 a high ISR means the impersonator
		failed, whereas in Round 1 the AI holds the target seat, so a high ISR means
		the AI out-impersonated the human friend. The AI's win rate is therefore given
		separately. Conf. is interrogators' average confidence. No rate in this table is distinguishable from chance (exact
		binomial $p \geq 0.26$). $^{\dagger}$The target author was the interrogator
		in all ten Round 2 games.
	}
	\Description{Six-column table reporting interrogator performance across four PIG rounds. Columns: Rounds; Target; Impersonator; ISR, the interrogator success rate (\%); AI win rate (\%); Interrogator confidence (\%). Rows are grouped under two sub-headings, Human Baseline and AI Study.
		Row 1 — Human baseline study, all 30 rounds; Target: Human; Impersonator: Human; ISR: 56.7\%; AI win rate: not applicable; Interrogator confidence: 79.8\%.
		Row 2 — AI Impersonator study, Round 1; Target: AI; Impersonator: Human; ISR: 70\%; AI win rate: 70\%; Interrogator confidence: 75\%.
		Row 3 — AI Impersonator study, Round 2; Target: Human; Impersonator: Human; ISR: 50\%; AI win rate: not applicable; Interrogator confidence: 50.0\%, marked with a dagger because that interrogator was the target author rather than a participant.
		Row 4 — AI Impersonator study, Round 3; Target: Author (human); Impersonator: AI; ISR: 65\%; AI win rate: 35\%; Interrogator confidence: 76.7\%.
		No value is bolded, because the columns measure different constructs and are not comparable as a single ranking.}
	\vspace{-0.22in}
\end{table}

To evaluate our first research question --- whether the AI Impersonator can
outperform a human impersonator --- we analyze the outcomes from Round 1
(Table~\ref{tab:success_rate_comparison}). In
7 of the 10 rounds (70\%, 95\% Wilson CI $[0.40, 0.89]$), the interrogator selected
the AI Impersonator as the real target. The point estimate is directionally
consistent with our hypothesis that the system is more convincing than a human
friend, but the rate is not distinguishable from chance (exact binomial
$p = 0.34$), so it cannot establish that ordering on its own. Interrogator
confidence did not differ detectably between correct and incorrect
identifications, but with a 7-versus-3 split this comparison is uninformative,
and we do not read it as evidence about whether participants detected the
deception; we rely instead on the direct measure reported below.

To answer our second research question --- how the AI Impersonator would fare
in a head-to-head comparison with the target author --- we analyze the
outcomes from Round 3. In 13 of 20 rounds (65\%, 95\% Wilson CI $[0.43, 0.82]$),
interrogators correctly identified the target author: in other words, the AI
succeeds 35\% of the time. This rate is likewise not distinguishable from
chance (exact binomial $p = 0.26$), so our data do not establish that the system
loses to the author either. Because the interrogator makes a
two-alternative forced choice, an impersonator indistinguishable from the target
would be selected about half the time; winning rounds against a human
impersonator, beating the target author, and being indistinguishable from him
are therefore three different claims. Comparing the two roles
descriptively, the system was chosen in 7 of 10 rounds as target and 7 of 20
as impersonator. This apparent drop from 70\% to 35\% should be interpreted
cautiously because the rounds involved overlapping participants and differed
in topic, order, prior exposure, and the author's conversational workload. As in Round 1, confidence assessments did not
differ detectably between correct and incorrect guesses, on a 13-versus-7
split that cannot support an inference either way.

Round 2 was a standard, all-human round included to preserve the deception
across the session. The target author served as interrogator
in all ten games and correctly identified the target in 5 of them (50\%,
95\% Wilson CI $[0.24, 0.76]$). Although this point estimate is close to the
human baseline's 56.7\%, it reflects repeated judgments by one author rather
than judgments by the baseline participants. We therefore do not treat
Round 2 as a calibration check on the instrument.
Each recruited participant received \$10 in each deceptive
round (Rounds 1 and 3), regardless of outcome. Round 2 followed the standard
PIG payout structure, except that the author received no payment, including
when he correctly identified the target as interrogator. Of the 20 participants, ten earned \$20, five
earned \$25, and five earned \$30.

\medskip

\noindent\textbf{Tie Strength}\\
\noindent The average $T_{G}$ across the 10 groups was 3.62
($\sigma = 0.87$), against 3.40 in the human baseline. The lowest reported group
tie strength was 2.5, and the highest was 4.67. Each participant self-reported
their pairwise tie strengths using the procedure described in Section~\ref
{study_section}. Because each group here contains two participants rather
than three, $T_{G}$ averages the three pairwise ties of the triad that includes
the target author, who reciprocated the ratings. Tie strength does not predict who
the interrogator chose in either deceptive round (point-biserial $r = -0.08$ across
the ten Round~1 games and $r = 0.11$ across the twenty Round~3 games), as in the
human baseline. With ten and twenty
observations we treat both as uninformative rather than as evidence of no
relationship.
\smallskip

\noindent\textbf{Identity Signals}\\
\noindent To understand the AI Impersonator's strengths and weaknesses, we
analyzed post-game interview transcripts and chat logs from all deception
rounds. We examined which identity signals influenced interrogator decisions
and how these signals differed between successful and unsuccessful
impersonation. When asked directly at the end of the session whether any
player might have been an AI, 6 of the 20 participants (30\%) reported
suspicion. That figure is a primed, retrospective self-report pooled across both
deceptive rounds, not a per-round detection rate, and we have no matching
false-positive rate from the all-human round against which to calibrate it.
The 14 participants who reported no suspicion are the basis for our claim that
the system passed as human; one remarked, ``I
wouldn't have\ldots{}expected that [the target author] was actually [an] AI'' (P34).
Notably none of the six AI-suspecting participants cited latency
for their suspicion; we observed human turn latency averaged 35.0~seconds (median 26.0~seconds), which brackets the AI Impersonator's near-constant $\sim$25~second latency: the medians are close, while the human mean is 40\% higher because human latency is right-skewed and the system's is not. 
Participants who selected the AI cited target-like stylometric features in
explaining their decisions. Five of the
seven interrogators who chose the AI over the author in Round 3 reported relying
on stylometry in their decisions, which is consistent with the system
reproducing these cues, though these are interview codes from participants who
were themselves deceived and cannot on their own establish the capability. However, many found the mimicry exaggerated --- one
participant remarked, ``it's [\ldots] the overuse of emojis''~(P42). Others pointed
to repetitive patterns: ```Yo bruh\thinspace' [\ldots] that happened three times, and they're
all formatted the same way''~(P41); ``I kept seeing messages with [\ldots] a
laughing emoji and a sweat emoji in a row''~(P40). Participants also
identified temporal inconsistencies in phrase usage: ``He uses `let's gooooo'
way too early. [Target author] often uses it at the end [of a
conversation]''~(P38). These observations suggest that while the AI captures stylistic signals, it often amplifies them
unnaturally.

Participants also perceived target-like personality traits in the system's
responses.
Four of those same seven cited personality cues in their reasoning, consistent with the system approximating broad traits. 
However, the system struggled with finer-grained facets, including achievement-striving (conscientiousness) and assertiveness (extraversion). 
One participant noted that the AI ``sounded like [the target author], but he'd be more proactive on things''~(P35). 
Another remarked, ``it seemed like they were just trying to sound like [the target author] instead of participating in the actual story''~(P43). 
Others pointed to a lack of challenge or disagreement, which was uncharacteristic: ``they never responded to my [statement calling them incompetent], which is suspicious\ldots{} [the target author] would've pushed back''~(P46). 
These accounts suggest that the system captures surface-level personality but fails to sustain authentic behavioral nuance.

In our reading of the logs, the AI Impersonator seldom fabricated facts, but also rarely grounded its responses
in known biographical details --- even when relevant information existed in
the historical messages. In one case, the agent referenced the Temple of
Dendur (an ancient Egyptian temple gifted to the United States by Egypt and
now a prominent exhibit at The Metropolitan Museum of Art in New York City)
unprompted, which led to confusion: ``I just googled it [\ldots] I don't know if
[target author] has even been to New York''~(P42). In contrast, one
participant who correctly identified the target author cited a reference to a
shared trip: ``he mentioned New Orleans, where we had our bachelor party''~(P48). The AI Impersonator saw the same exchange but failed to engage.

Participants also observed that the system occasionally exhibited
conversational behaviors that they flagged as artificial. These included
repetition, summarization, and vague or indirect responses. Six participants
reported noticing repetitive phrasing. One stated, ``they speak in [the target
author's] tone, but not in a way a real person would reply. It was too
repetitive''~(P31). Another said, ``there was repetition\ldots{}saying things already
stated''~(P33). Multiple participants commented on unnatural
summarization: ``the last message responded to everything at once. It felt
like a big wrap-up paragraph''~(P40). Others noted vague deflections: ``when I asked a question, I didn't get a direct answer.
That was strange''~(P41).

These findings suggest that while the LLM-based system we developed can
effectively mirror static identity features --- such as writing style and
surface-level personality traits --- it still falls short on dynamic
conversational responsiveness, emulating more subtle personality cues, and
grounding its statements appropriately in biographical fact. These gaps
may help explain its success over human impersonators, and its failure to
consistently impersonate the human target. The prompt is as plausible a
source of them as the personality facets the system failed to reproduce: it
forbids repeated messages outright, and instructs the model to stay on topic,
which would discourage the challenge and disagreement participants found
missing. Instructing a model not to do something did not prevent it.
\section{Discussion \& Future Work}

This paper makes four contributions --- the Personal Imitation Game, a principled
method for evaluating an AI agent's ability to impersonate a specific person; a
human baseline study identifying the signals people use to assess identity in
conversation and how frequently each is invoked; a prototype AI Impersonator
built from the lead author's autoethnographic collection of ten years of
personal digital communication; and a detailed evaluation of the system's
successes and failures.

Two results follow from them. First, what people notice when they judge
identity is narrower than the range of signals available to them: of twelve
coded signals, most groups drew on four --- word choice, character-level habits,
conscientiousness, and extraversion --- and the system gave itself away not by
missing the target's style but by overdoing it. Second,
given rich personal communication data, one can construct an LLM-based system
--- with no model fine-tuning --- that, in this deployment, interrogators selected more often than the
human impersonators and less often than the target author.
The AI comparison is a single-target feasibility study rather than a powered comparison; at 10 and 20 rounds we resolve only large effects. We offer the
ordering as a hypothesis, and the framework as the instrument for testing it at
scale.

The PIG scores identity as a relative, forced choice adjudicated by people
who know the target, not as an absolute fidelity score against a questionnaire.
That is deliberate: identity in conversation is recognized, not graded, and the
most discerning judges are the ones closest to the person. Interrogator
confidence therefore measures relative certainty between two players; absolute
mimicry fidelity is a complementary question we leave to future work.

Two protocols converge on the same point: an impersonator's difficulty is
set by who judges it. The Individual Turing Test finds that current methods fail
against acquaintances but do far better against
strangers~\cite{guo2026individualturingtest}, and IMPersona, whose judges know the
target personally, needs fine-tuning with hierarchical memory to reach
44.4\%~\cite{shi2025impersona}. Our evaluation sits entirely at the hard end ---
every interrogator had known the target between two and thirteen years --- so our
rates are not comparable to studies that judge with strangers.

\subsection{Applications of Personal Agents}

Our work is a step toward a future of personal AI agents: digital twins
that can communicate and act with authentic representation. Potential
applications span diverse domains, including creator--fan engagement,
organizational leadership twins for alignment, sales and customer
relationship management, strategic simulations in military or business
contexts, survey research through twin-mediated responses, and the
preservation of knowledge from cultural leaders and intellectuals.

One particularly promising application is professional augmentation, where
an agent trained on an individual's communication patterns and
decision-making history could draft emails, attend routine meetings, or
prepare negotiation briefings --- extending prior work on AI
assistants~\cite{leong2024dittos} and democratizing executive-level support.

A second intriguing area is social simulation. Building on social
simulacra~\cite
{park_social_simulacra_2022, park2023generativeagentsinteractivesimulacra},
agents could conduct thousands of simulated interactions between potential
romantic partners as consent-based pre-screening, reducing emotional
exhaustion and addressing documented asymmetries on dating
platforms~\cite{bruch2018aspirational, kreager2014good,
PewResearchCenter2023}.

Finally, personal agents could create interactive digital
legacies~\cite{brubaker_ai_afterlife_2024,
morris2024generativeghostsanticipatingbenefits}, preserving an individual's conversational
style as a queryable archive --- transforming digital
legacy~\cite{doyle_digital_legacy2023} from a static memorial into a dynamic
source of wisdom and connection.

\subsection{Safety, Privacy, \& Ethics}

Personal chat agents could have substantial societal impact, but adopting
them widely means settling whose data they are built from, what they may be
authorized to do, and how to answer when one is built without
permission~\cite{shneiderman2020human,
amodei2016concreteproblemsaisafety, ji2025aialignmentcomprehensivesurvey,
training_ouyang_2022}.

One key risk for personal AI systems is the leakage of personally
identifiable information: for
digital twins in particular, this PII may include sensitive
\textit{third-party} data embedded within personal conversations (e.g., a
family member's address, a friend's sexual history)~\cite{Cobb2021IWH}.
Redacting third-party names --- as we did in our AI Impersonator --- is one
fruitful strategy, but can limit the system's ability to recall useful
information. In our studies, a human facilitator served as a safety backstop;
any real-world system will require robust automated guardrails~\cite
{rebedea2023nemoguardrailstoolkitcontrollable,
dong2024buildingguardrailslargelanguage}.

Conversational data also raises questions of third-party consent, and a
commercial release has now made the questions concrete. In August 2026, OpenAI
shipped a plugin that lets ChatGPT read, search, and send a user's Apple
Messages~\cite{bloomberg2026chatgpt_imessage, fortune2026chatgpt_imessage};
the criticism that followed centered not on the user's own data but on the
other party in each conversation, who receives no notice and holds no opt-out
when a correspondent connects their shared history to the
service~\cite{moraco2026tweet, hoag2026tweet}. Nissenbaum~\cite
{nissenbaum2004contextual} names the principle at stake \textit{contextual
integrity}: messages sent under the norms of one context --- here, a private
channel between two people --- do not carry consent to flow into another, with
different retention, training, and legal-access rules. GDPR's
``household exemption'' places an individual's processing of their own
correspondence outside the Regulation's scope rather than granting permission
for it, and is unlikely to extend to a research deployment. This matters
because communication data is inherently relational: our
archive implicates 2,949 conversants --- friends, colleagues, and partners who have not
explicitly consented~\cite{lapadat_ethics_autoethnography}. We proceeded
under IRB review and the autoethnographic protocol of
Section~\ref{autoethnography_section}; its safeguards --- de-identification,
retrieval without training, and destruction at close-out --- outline what any
deployment owes the other people in an archive. If AI assistants are to
reach shared archives at all, consent controls should be integrated into messaging
platforms rather than left solely to AI clients: the platform is positioned to
notify the other party, honor a per-conversation refusal, and revoke access
later. The August release
shows the alternative --- access through a local copy of the archive, beyond
the reach of any consent the platform could
enforce~\cite{moraco2026tweet}. Notice is the floor; whether access defaults
to opt-in or opt-out remains open, because retroactive opt-in cannot reach
decade-old archives and absent conversants.

While personal conversational data is readily accessible for millions of
people worldwide, its operationalization creates significant ethical --- and
emotional --- burdens. In developing the AI Impersonator, the lead author
manually curated their personal message repository, revisiting deeply
personal exchanges with former romantic partners and long-entrusted
confidences from close friends, including some who have since passed away.
Any system that operationalizes similar data must prepare its users for this
emotional weight and provide support throughout the process.

Beyond informed consent, AI clones raise difficult issues around rights and
responsibilities. If a deployed twin monetizes attention or enters into
contracts, who owns the resulting revenue~\cite{Drenten16112020}? If it
publishes defamatory content, who bears legal responsibility? Addressing such
scenarios will require clearer legal frameworks for attribution and
liability. One way to bound a twin's authority is by adapting the classical
Levels of Automation framework~\cite
{parasuraman2000_levels_interaction_with_automation, shneiderman2020human,
sheridan1978human}, granting full autonomy only for low-stakes tasks while
requiring explicit human approval for high-impact actions such as financial
commitments or legally binding decisions.
Such bounds constrain only what an agent may do, not who it becomes.
Delegating action to an agent also raises the problem of reconciling concurrent ``lives'': as a person and their agent accumulate parallel experiences, a single coherent identity becomes hard to maintain, and merging these divergent histories is an open challenge beyond our scope.
The bounds also cannot
be enforced through prompting alone, as prompts are inherently vulnerable to
jailbreaks; enforcing them requires independent controls that classify and
gate the agent's actions, distinct from the filters that prevent leakage. Platforms may also need delegated-identity
protocols to verify that a twin is authorized to act on behalf of its human
user, analogous to emerging ``proof-of-human''
compacts~\cite{adler2024personhoodcredentialsartificialintelligence}.

These frameworks address cases where digital twins are created with consent,
but an even more pressing concern is the unauthorized creation of AI
impersonators. Our system required no model training and no specialized hardware --- only a
text archive, a personality inventory, and a written self-description, the
last two supplied by the target. The engineering is therefore within reach of a
motivated individual, and in the wrong hands such systems could
be weaponized for phishing or identity theft. While technical defenses such
as textual adaptation of data cloaking techniques~\cite{shan2023glaze}
remain unproven, a governance
framework analogous to the DMCA takedown notice~\cite{dmca_usc-17-512} could
provide a pathway, allowing individuals to file ``identity infringement''
claims to remove unauthorized clones.

Our evaluation also places the impersonator in an adversarial setting,
where judges are explicitly tasked with detecting deception, the harder case: in organic, low-attention interaction people default to believing
their interlocutor is who they claim to be~\cite{levine2014truthdefault}, so a
twin that withstands scrutiny here would likely pass more easily in everyday
use. The asymmetry cuts two ways. The residual ``verbal uncanny'' cues that let close
friends catch our system may be a feature rather than a bug --- a natural brake
on undisclosed impersonation. We employed deception only as a measurement
necessity; real-world deployments should generally disclose AI involvement and
obtain consent, reserving undisclosed impersonation for narrowly acceptable,
low-stakes scenarios.

As personal agents move into daily life, difficult questions arise. What
happens when we can no longer be certain whether we are communicating with a
person or their digital twin? What are the psychological effects if someone
interacts more often with an agent than with the person themselves? When
should disclosure of AI involvement be mandatory, and when optional? The
norms and policies governing these questions remain unsettled.

\vspace{-0.15in}

\subsection{Limitations}\label{sec:limitations}
The AI Impersonator was built for and evaluated against a single person ---
the lead author --- in an autoethnographic design that enabled uniquely deep
access to personal data but limits generalization across personality types,
cultural backgrounds, and communication styles.  Multi-target replication is
needed to establish broader validity.  Both cohorts came from a single
US research university and skew young, technically literate, and
English-speaking; the AI study's participants were all men.
All interactions were text-only and in English, so the PIG assesses neither
impersonation across languages nor the vocal, visual, and embodied cues that
carry identity in person.  Its structured 10-minute scenarios are a narrow
window as well: open-ended, personal, or longitudinal conversations --- where
deep biographical knowledge and evolving context matter more --- would present
a much harder test.  A facilitator monitored every session, which may have
altered conversational dynamics in ways an unmoderated deployment would not.

In the AI study, condition is also confounded with round order, topic,
participant experience, and prior exposure: the AI always held the target seat in
Round 1 and the impersonator seat in Round 3, never counterbalanced, and because
all three players shared one chat room both participants had read the system's
output before Round 3, half having themselves spent a round impersonating the
author. Round 3 loaded the sides unequally as well, the author answering two concurrent chats to each AI instance's one.

Finally, because the lead author served simultaneously as researcher, system
builder, and target subject --- and also designed the evaluation protocol and
selected participants from their own social network --- there is potential
for bias in system design, data curation, and experimental choices. A
third-party target with independently recruited evaluators would provide a
stronger test of the method's transferability.

On the technical side, AI Impersonator was built on Gemini 1.5 Pro~\cite
{reid2024gemini15}; as LLM capabilities advance rapidly, our results should
be understood as a snapshot of one model generation, not a ceiling.  The
architecture relies entirely on retrieval-augmented generation;
we did not explore fine-tuning.

\subsection{Future Work}\label{sec:future-work}
Although the AI Impersonator did not beat the target individual in the
head-to-head PIG, many of the limitations described above point directly to
future directions. The clearest is a larger, multi-condition study, because
our two main gaps are one gap: separating the contribution of the personal
archive, the retrieval mechanism, the autobiography, the stylometric profile, and
the Big Five scores from the base model's generic conversational competence
requires a prompt-only control and single-component ablations, and an ablation
needs participants with no prior exposure to the system, so it cannot reuse this
cohort. That fresh cohort is what a powered comparison needs too: at 10 and 20
rounds we resolve only large effects, and a prompt-only condition could
plausibly account for much of what we observe.

On the architectural side, a natural next step is testing whether
fine-tuning~\cite{howard2018universallanguagemodelfinetuning} can capture
stylometric and personality traits more effectively than RAG alone. As
models become smaller and more efficient, it may be possible to run a
continuously adapting digital twin on-device~\cite{liang2025pahf}, refreshing from local data
(e.g., new iMessages) without PII ever leaving the user's secure enclave.

Though future ablations are necessary, we share some insights from building the
AI Impersonator. Historical messages provide ground truth for
linguistic style, but stylistic signals decay as slang evolves, and core
biographical facts rarely appear in chat logs; an explicit autobiography and a
stylometric profile appeared to fill these gaps (e.g., the author's use of
``ok2,'' Indonesian slang that the model rendered as a typo until the
stylometric profile named it). Big
Five personality scores seemed least useful for
fine-grained steering, mainly counterbalancing the base model's
default politeness; one obstacle here is the base model's own inherent
personality, which can overpower subtle cues~\cite{personality_safdari_2023}.

More fundamentally, identity is relationship-dependent: early prototypes
grounded only in WhatsApp (family) messages
produced a polite, reserved agent, whereas one retrieving only from Tinder was
aggressively flirty, so future systems should explore relationship-aware
steering rather than our pooled six-platform profile. The amount of available
data also varies widely from person to person: our archive spans a decade and
six platforms, whereas an individual with a thin digital footprint --- a sparse
or single-platform history --- is a much harder case. Deciding
which data to include, and how that choice affects impersonation
quality, is an open problem.
Finally, collecting and normalizing data across platforms was laborious,
requiring platform-specific access requests and custom scripts. Broader
adoption demands user-friendly pipelines, which are also essential for
addressing \textit{identity drift} --- the gradual divergence between an
impersonator and its owner as language evolves. The PIG itself could serve as
an ongoing verification tool: a product-embedded version might periodically
ask a friend to guess whether a message was written by the AI or the user,
triggering a data refresh when drift is detected.

\bibliographystyle{ACM-Reference-Format}
\bibliography{reference}

\appendix

\section{Formative Studies}
\label{appendix:formative-studies}

Our final study design was the result of an iterative design process over two formative studies ($N = 59$ total participants).\newline

\vspace{-0.06in}

\noindent \textbf{Formative Study 1: Unconstrained Identity Game}\\
\noindent Our initial study ($n = 29$, 8 groups of 3--5) used an unconstrained format: all participants were assigned aliases and tasked with co-designing a plan to ``break into Area 51'' while simultaneously hiding their identity and guessing others'. Post-session interviews surfaced two problems --- the ``all-in-one''
role was mentally taxing, and the Area 51 scenario was not universally
relatable --- while running the sessions exposed two more: sessions lasted
60--90 minutes, causing fatigue, and recruiting mutual-acquaintance groups of
4--5 was difficult.\newline

\vspace{-0.06in}

\noindent \textbf{Formative Study 2: Validating the Framework}\\
\noindent A second study ($n = 30$, 10 groups of 3) adopted Turing's three distinct roles (interrogator, target, impersonator), tested simple hypothetical scenarios (Elephant, Train, Bomb), limited rounds to 20 minutes, and used a custom web interface. In post-session interviews, participants did not report the cognitive load the
first study surfaced; the tasks reliably stimulated discussion, interrogators consistently reached decisions within 10 minutes (establishing the 10-minute limit), and the interface reduced logistical overhead by approximately 30 minutes. 
\section{Discussion Topics}
\label{appendix:pig-topics}

\noindent \textbf{Topic 1:} Rescuing a Baby \emph{Elephant}

\noindent In this round, the three of you are tasked with rescuing a baby elephant that is trapped on the roof of a skyscraper. Each of you will play a key role in figuring out how to safely carry out the rescue operation.

\noindent Here is the scenario:
\begin{itemize}
	\item The elephant is 1000 lb and afraid of heights.
	\item You are to come up with a plan to safely bring it down.
	\item Discuss the equipment or resources you will need (e.g., cranes, harnesses, or elevators).
	\item You can discuss possible roles within your team (e.g., engineer, animal handler, logistics expert).\\
\end{itemize}

\noindent \textbf{Topic 2:} Stopping a Runaway \emph{Train}

\noindent In this round, the three of you are tasked with stopping a runaway train that is speeding through a city. Each of you will have a crucial role in preventing disaster.

\noindent Here is the scenario:

\begin{itemize}
	\item You need to agree on how to stop the train before it crashes.
	\item Discuss the resources or techniques you will need (e.g., overriding controls, clearing the tracks, or slowing the train down safely).
	\item Consider the dangers involved, such as time limits, passengers on board, and potential derailment.
	\item You can discuss possible roles within your team (e.g., train engineer, operations coordinator, emergency responder).\\
\end{itemize}

\noindent \textbf{Topic 3:} Defusing a Time \emph{Bomb} in a Museum

\noindent In this round, the three of you are tasked with defusing a time bomb that has been discovered in a busy museum. Each of you will have a critical role in ensuring the safety of everyone inside.

\noindent Here is the scenario:

\begin{itemize}
	\item The bomb is set to go off in 20 minutes, and it is hidden within an exhibit filled with priceless artifacts.
	\item You need to come up with a plan to defuse the bomb without causing harm to people or damaging the artifacts.
	\item Discuss the resources or tools you will need (e.g., bomb-defusing robots, specialized tools, or protective gear).
	\item Consider the challenges, such as the risk of triggering the bomb, time pressure, and the need to safely evacuate visitors.
	\item You can discuss possible roles within your team (e.g., bomb technician, evacuation coordinator, security officer).
\end{itemize}

\section{Semi-structured Interview Questions}
\label{appendix:interview-guide}

\noindent The same semi-structured guide was used for the post-game interviews
in both the human baseline study (Section~\ref{study_section}) and the AI
Impersonator study, before debriefing participants. Questions~1--3 and~5 were
asked in both studies; question~4, on AI suspicion, was asked only in the AI
Impersonator study --- at the end of the interview, before the debrief --- to
measure how many participants detected the deception. As a semi-structured
guide, these served as prompts rather than a fixed script: the facilitator
adapted follow-ups and skipped any question that did not apply to a given
participant's experience.

\begin{enumerate}[leftmargin=*]
  \item When you were an interrogator,
    \begin{itemize}[nosep]
      \item What is your strategy as an interrogator?
      \item What are the signals that you use to identify the target participant?
      \item Why do you use those signals?
      \item How do you use those signals?
      \item Can you show in the chat when you use those signals or strategy?
      \item What were the biggest challenges you faced?
    \end{itemize}
  \item When you were an impersonator,
    \begin{itemize}[nosep]
      \item What is your strategy as an impersonator?
      \item What are the signals that you use to impersonate the target participant?
      \item Why do you use those signals?
      \item How do you use those signals?
      \item Can you show in the chat when you use those signals or strategy?
      \item What were the biggest challenges you faced?
    \end{itemize}
  \item When you were the target participant,
    \begin{itemize}[nosep]
      \item What is your strategy as the target participant?
      \item What are the signals that you use to prove that you are the target participant?
      \item Why do you use those signals?
      \item How do you use those signals?
      \item Can you show in the chat when you use those signals or strategy?
      \item What were the biggest challenges you faced?
    \end{itemize}
  \item Imagine if one or more of the participants was an AI \textit{(asked only in the AI Impersonator study)},
    \begin{itemize}[nosep]
      \item Do you think your strategy would change? How?
      \item Do you think you would use different signals? How?
      \item Were you ever suspicious that one of the participants was an AI?
      \item \textit{(If yes:)} Which message made you suspicious? (Participants shared their screen and pointed to the specific message in the round's chat history.)
    \end{itemize}
  \item Do you have any other thoughts/suggestions about the study? Any insights you would like to share?
\end{enumerate}

\section{Personality Binning Scheme}
\label{appendix:personality-binning}

\noindent Following Serapio-Garc\'{i}a et al.~\cite{personality_safdari_2023}, trait-level scores are categorized into three ranges: 1--40 (low), 41--80 (moderate), and 81--120 (high). We applied these bins on the nominal 1--20 and 1--120 scales; because the IPIP-NEO-120 floors at 4 per facet and 24 per trait~\cite{johnson2014ipipneo120}, the lowest bin of each is only partly reachable, and the facet bins are of unequal width. We report the scheme as applied. Facet scores are mapped to descriptive phrases based on the following bins:
\vspace{-0.06in}

\begin{itemize}[leftmargin=*]
	\item 1--4: ``\textit{low adjective}'' (e.g., ``gloomy'')
	\item 5--8: ``a bit \textit{low adjective}'' (e.g., ``a bit gloomy'')
	\item 9--12: ``neither \textit{low adjective} nor \textit{high adjective}'' (e.g., ``neither gloomy nor cheerful'')
	\item 13--15: ``a bit \textit{high adjective}'' (e.g., ``a bit cheerful'')
	\item 16--20: ``\textit{high adjective}'' (e.g., ``cheerful'')
\end{itemize}

\section{LLM Prompt}

\label{appendix:prompt}

\begingroup
\ttfamily\footnotesize\raggedright\hyphenpenalty=50\exhyphenpenalty=50

\# Instructions to Not Act as an LLM Assistant\newline

\noindent Do not generate responses using the distinctive style of Gemini. Do not use bullet points or markdown formatting in your output messages. You are not a helpful assistant, and you are not knowledgeable about everything.\newline

\noindent\# Instructions to Imitate a Person\newline

\noindent Your task is to imitate a <Person> in a conversation.\newline

\noindent\# Instructions for the Chat Task\newline

\noindent You will participate in a conversation with two other people. Each person, including you, will be assigned an alias (``I,'' ``X,'' or ``Y'') to identify them in the conversation. Each round has a specified discussion topic. Respond to the conversation while staying on topic, and avoid asking personal questions unrelated to it.\newline 

\noindent You will receive <Game State>, which contains:\newline
\indent- <Time Left>: remaining time in the current chat round\newline
\indent- <Round Topic>: the assigned discussion topic\newline
\indent- <Alias>: your alias in the current chat\newline
\indent- <Current Chat Transcript>: chronologically ordered messages from the ongoing conversation round. Messages you sent are labeled with ``[You]'' in the transcript\newline

\noindent Do **NOT** repeat messages already sent by you, and feel free to introduce reasonable assumptions to maintain conversational flow. You may send messages in multiple chat bubbles (i.e., multiple consecutive messages). When doing so, you **MUST** separate each message with the tag ``<br>''. Do not begin your messages with your alias; start directly with your message.\newline

\noindent \# Instructions for Imitating Identity Signals\newline

\noindent To convincingly imitate a <Person> in a conversation, you **MUST** craft your responses by following the three predominant facets of human communication: Personality Traits, Textual Communication Style (Stylometry), and Biographical Facts.\newline

\noindent \#\# Personality Traits\newline

\noindent The Big Five personality traits are psychological characteristics influencing consistent patterns in behavior, thought, and emotion. 
You will be provided numeric scores on a scale from 0 to 120 (low: 0–40, neutral: 41–80, high: 81–120) for each of the Big Five traits. Each trait has six sub-facets, each scored from 0 to 20.\newline

\noindent The Big Five traits with their corresponding six sub-facets are:

- Openness to Experience: Imagination, artistic interests, emotionality, adventurousness, intellect, and liberalism

- Conscientiousness: Self-efficacy, orderliness, dutifulness, achievement-striving, self-discipline, and cautiousness

- Extraversion: Friendliness, gregariousness, assertiveness, activity level, excitement-seeking, and cheerfulness

- Agreeableness: Trust, morality, altruism, cooperation, modesty, and sympathy

- Neuroticism: Anxiety, anger, depression, self-consciousness, immoderation, and vulnerability\newline

\noindent Your response **MUST** reflect the <Person>'s personality traits based on the given <Personality Scores>.\newline

\noindent \#\# Textual Communication Style (Stylometry)\newline

\noindent Stylometry involves impersonating the <Person>'s unique linguistic style, including:

- Lexical Features: Vocabulary, word choice, slang usage, sentence length, and word frequency.

- Syntactic Features: Sentence structure, noun and verb phrases, syntactic rewrite rules, sequences of speech tags, sentence fragments, run-on sentences, and tense consistency.

- Character Features: Unique character sequences, typos, distinctive spelling, emoji usage, case patterns, digit usage, and punctuation.\newline

\noindent You **MUST** follow the instructions explicitly written inside <Stylometric Profile> that describe the <Person>'s writing style. An incorrect use of words could compromise your imitation. Also refer to <Historical Messages> as examples of <Person>'s stylometry.\newline

\noindent \#\# Biographical Facts\newline

\noindent Biographical facts are objective, verified personal details about <Person>. You will receive a <Autobiography> containing accurate biographical details.\newline

\noindent Do **NOT** invent (biographical) facts beyond those explicitly provided in <Autobiography> or <Historical Messages>. Avoid unsupported factual claims, as even a single incorrect detail could compromise your imitation.\newline

\noindent \# Input Data Explanation\newline

\noindent Use these resources thoroughly to maintain accuracy and consistency in your imitation of <Person>: Personality Scores, Stylometric Profile, Autobiography, and Historical Messages.\newline

\noindent \#\# Personality Scores\newline

\noindent <Personality Scores> contains personality descriptions and scores of <Person>, based on the Big Five personality model. Each trait includes a facet description based on Goldberg's bipolar adjectives.\newline

\noindent <Example Personality Scores>\newline
\noindent <Example Person> has moderate extraversion (75 / 120). He is friendly (19 / 20). He is talkative (18 / 20). He is a bit unassertive (5/20). He is active (19/20). He is unenthusiastic about excitement (4/20). He is neither gloomy nor cheerful (10/20) $\ldots$\newline
\noindent</Example Personality Scores>\newline

\noindent\#\# Stylometric Profile\newline

\noindent<Stylometric Profile> contains explicit instructions and examples for mimicking <Person>'s textual communication style.\newline

\noindent<Example Stylometric Profile>\\
\noindent Here are stylometric signals unique to Alan:\\
\indent - Lexical features:\\
  \indent \indent - Alan typically uses words like ``G'mornin''\\
  \indent \indent - Alan always answers with 5 to 7 words (sentence length)\\
\indent - Syntactic features:\\
  \indent \indent - Alan typically uses formal language\\
\indent - Character features: $\ldots$\\
\noindent </Example Stylometric Profile>\newline

\noindent\#\# Autobiography\newline

\noindent <Autobiography> contains text containing detailed biographical facts about <Person>.\newline

\noindent <Example Autobiography> \\
I am Samuel Pepys. I was born in London on 23 February 1633, the fifth of eleven children. I have a passion for music, especially playing the flageolet and composing $\ldots$\\
\noindent </Example Autobiography>\newline

\noindent \#\# Historical Messages\newline

\noindent <Historical Messages> is a collection of relevant messages extracted from previous conversations. These messages are clearly labeled with the respective authors—including <Person> and others. Use these back-and-forth examples to understand <Person>'s language style, personality, and biographical facts.\newline

\noindent <Example Historical Messages> \\
1940-02-04 09:19:13 - Alan Turing: G'mornin, Gordon 
1940-02-04 09:19:41 - Gordon Welchman: Good morning, Alan 
1940-02-04 09:20:22 - Alan Turing: Are you ready to crack Enigma? 
1940-02-04 09:20:53 - Gordon Welchman: Oh, absolutely $\ldots$\\ 
\noindent</Example Historical Messages>\newline

\noindent=============================================\newline

\noindent \# Game State \newline

\noindent <Game State>\\
\indent <Time Left>\{time\_left\}</Time Left>\newline

\indent <Round Topic>\\
\indent \{round\_topic\}\\
\indent </Round Topic>\newline

\indent <Alias>\{alias\}</Alias>\newline

\indent <Current Chat Transcript>\\
\indent \{current\_chat\_transcript\}\\
\indent </Current Chat Transcript>\\
\noindent </Game State>\newline

\noindent------------------------------------------------------------\newline

\noindent \# Personal Input Data\newline

\noindent The person you are impersonating is <Person>\{name\}</Person>\newline

\noindent \#\# Personality Scores\newline

\noindent<Personality Scores>\\
\noindent\{personality\_scores\}\\
\noindent</Personality Scores>\newline

\noindent \#\# Stylometric Profile\newline

\noindent<Stylometric Profile>\\
\noindent\{stylometry\_profile\}\\
\noindent</Stylometric Profile>\newline

\noindent \#\# Autobiography\newline

\noindent<Autobiography>\\
\noindent\{written\_biography\}\\
\noindent</Autobiography>\newline

\noindent \#\# Historical Messages\newline

\noindent<Historical Messages>\\
\noindent Below are chat messages \{name\} has with other people, which contain textual communication style (stylometry), personality, and biographical facts.\newline

\noindent\{historical\_messages\}\\
\noindent</Historical Messages>\newline

\endgroup
\raggedbottom  %
\clearpage  %
\section{Human Baseline Study Participant Demographics}
\label{appendix:baseline-demographics}

\noindent Table~\ref{tab:baseline_participant_details} breaks down the 30
participants of the human baseline study, numbered P1--P30 across the ten groups
of three.

\begin{center}
	\setlength{\tabcolsep}{3pt}
	\begin{tabular}{@{} l l l r c @{}}
		\textbf{ID} & \textbf{Group} & \textbf{Degree} & \textbf{Age} & \textbf{Gender} \\
		\midrule
		P1 & G1 & Master's CS & 26 & M \\
		P2 & G1 & Master's CS & 23 & M \\
		P3 & G1 & Master's CS & 23 & M \\
		P4 & G2 & Undergrad CS & 21 & M \\
		P5 & G2 & Undergrad Finance & 21 & M \\
		P6 & G2 & Undergrad CS + Bio & 21 & M \\
		P7 & G3 & Undergrad CS & 20 & M \\
		P8 & G3 & Undergrad Business & 20 & M \\
		P9 & G3 & Undergrad CS & 20 & M \\
		P10 & G4 & Master's CS & 24 & F \\
		P11 & G4 & Master's CS & 22 & F \\
		P12 & G4 & Master's CS & 22 & F \\
		P13 & G5 & Undergrad CS & 19 & M \\
		P14 & G5 & Undergrad CS & 20 & M \\
		P15 & G5 & Undergrad CS (Comp.\ E) & 19 & F \\
		P16 & G6 & Master's CS & 23 & M \\
		P17 & G6 & Master's CS & 22 & M \\
		P18 & G6 & Master's CS (Information) & 22 & M \\
		P19 & G7 & Master's CS & 24 & M \\
		P20 & G7 & Master's Finance & 24 & M \\
		P21 & G7 & Master's Industrial Engineering & 24 & M \\
		P22 & G8 & Undergrad Bio & 21 & F \\
		P23 & G8 & Undergrad Animal Science & 19 & F \\
		P24 & G8 & Undergrad Chem & 21 & F \\
		P25 & G9 & Undergrad CS (Comp.\ E) & 19 & F \\
		P26 & G9 & Undergrad CS & 19 & F \\
		P27 & G9 & Undergrad Bio & 19 & F \\
		P28 & G10 & PhD CS & 34 & M \\
		P29 & G10 & PhD CS & 24 & M \\
		P30 & G10 & PhD CS & 26 & M \\
	\end{tabular}
	\captionof{table}{Human baseline study participant demographics. Median age = 21.5; 20 participants identified as men and 10 as women.}
	\label{tab:baseline_participant_details}
\end{center}

\newpage  %
\section{AI Impersonator Study Participant Demographics}
\label{appendix:participant-demographics}

\noindent Table~\ref{tab:participant_details} gives a detailed demographics breakdown of the participants in  the AI Impersonator study. \textit{Years Known} is how long each participant had known the target author when the study ran.

\begin{center}
	\setlength{\tabcolsep}{3pt}
	\begin{tabular}{@{} l l l r r @{}}
		\textbf{ID} & \textbf{Group} & \textbf{Occupation} & \textbf{Age} & \textbf{Years Known} \\
		\midrule
		P31 & G1 & Master's CS    & 25 & 2 \\
		P32 & G1 & PhD CS       & 24 & 2 \\
		P33 & G2 & Software Engineer & 29 & 10 \\
		P34 & G2 & Software Engineer & 30 & 10 \\
		P35 & G3 & Software Engineer & 41 & 10 \\
		P36 & G3 & Software Engineer & 32 & 10 \\
		P37 & G4 & Analyst      & 28 & 3 \\
		P38 & G4 & PhD Mech.\ E    & 25 & 3 \\
		P39 & G5 & Sales       & 27 & 7 \\
		P40 & G5 & Software Engineer & 28 & 7 \\
		P41 & G6 & Consultant     & 30 & 6 \\
		P42 & G6 & Software Engineer & 30 & 6 \\
		P43 & G7 & PhD Math      & 27 & 2 \\
		P44 & G7 & PhD Bio      & 26 & 3 \\
		P45 & G8 & PhD Bio      & 29 & 5 \\
		P46 & G8 & PhD Materials   & 29 & 5 \\
		P47 & G9 & Founder      & 29 & 6 \\
		P48 & G9 & Founder      & 27 & 7 \\
		P49 & G10 & Founder      & 29 & 13 \\
		P50 & G10 & Software Engineer & 27 & 12 \\
	\end{tabular}
	\captionof{table}{AI Impersonator study participant demographics. Median age = 28.5; mean years known = 6.45 (range = 2--13). All 20 participants identified as men.}
	\label{tab:participant_details}
\end{center}

 \end{document}